\documentclass[article, onecolumn, 11pt]{IEEEtran}

\usepackage[english]{babel}
\usepackage[utf8]{inputenc}
\usepackage{amsmath}
\usepackage{graphicx}
\usepackage[colorinlistoftodos]{todonotes}
\usepackage{hyperref}
\usepackage{scrextend}

\usepackage{multirow}

\usepackage{bm}

\usepackage{graphicx}
\usepackage{color}
\usepackage{placeins}
\usepackage{float}
\usepackage{tabularx,colortbl}

\usepackage{amsfonts}
\usepackage{subfigure}
\usepackage{bm}
\usepackage{url}
\usepackage{multirow}
\usepackage{caption}

\usepackage{amssymb}

\usepackage{algorithm}
\usepackage[noend]{algpseudocode}
\makeatletter
\def\BState{\State\hskip-\ALG@thistlm}
\makeatother

\usepackage{amssymb}

\newtheorem{thProth}{Theorem}
\newtheorem{defin}{Definition}

\ifCLASSINFOpdf

\else

\fi

\begin{document}
\title{Number Theoretic Transforms for Secure Signal Processing\thanks{This work was partially funded by the Spanish Ministry of Economy and Competitiveness and the ERDF under projects TACTICA, COMPASS (TEC2013-47020-C2-1-R) and COMONSENS (TEC2015-69648-REDC), and the FPI grant (BES-2014-069018), by the Galician Regional Government and ERDF under projects GRC2013/009 and AtlantTIC, and by the EU H2020 Framework Programme under project WITDOM (proj. no. 644371).}}

\author{Alberto~Pedrouzo-Ulloa,~\IEEEmembership{Student Member,~IEEE,}
  Juan~Ram\'on~Troncoso-Pastoriza,~\IEEEmembership{Member,~IEEE,}
  and~Fernando~P\'erez-Gonz\'alez,~\IEEEmembership{Fellow,~IEEE}
  \thanks{A. Pedrouzo-Ulloa, J.R. Troncoso-Pastoriza and F. P\'erez-Gonz\'alez are with the Department
    of Signal Theory and Communications of the University of Vigo, Vigo, 36310 Spain e-mail:\{apedrouzo,troncoso,fperez\}@gts.uvigo.es.}}

\maketitle

\begin{abstract}
  Multimedia contents are inherently sensitive signals that must be protected whenever they are outsourced to an untrusted environment. This problem becomes a challenge when the untrusted environment must perform some processing on the sensitive signals; a paradigmatic example is Cloud-based signal processing services. Approaches based on Secure Signal Processing (SSP) address this challenge by proposing novel mechanisms for signal processing in the encrypted domain and interactive secure protocols to achieve the goal of protecting signals without disclosing the sensitive information they convey.
  
  This work presents a novel and comprehensive set of approaches and primitives to efficiently process signals in an encrypted form, by using Number Theoretic Transforms (NTTs) in innovative ways. This usage of NTTs paired with appropriate signal pre- and post-coding enables a whole range of easily composable signal processing operations comprising, among others, filtering, generalized convolutions, matrix-based processing or error correcting codes. Our main focus is on unattended processing, in which no interaction from the client is needed; for implementation purposes, efficient lattice-based somewhat homomorphic cryptosystems are used. We exemplify these approaches and evaluate their performance and accuracy, proving that the proposed framework opens up a wide variety of new applications for secured outsourced-processing of multimedia contents. 
\end{abstract}


\IEEEpeerreviewmaketitle
\section{Introduction}
\label{sec:intro}
\IEEEPARstart{S}{ignal} processing is present in virtually any every-day digital appliances and applications, from voice processing for telephony to complex image processing in rendering 3D-movies, covering also biometric processing of faces, fingerprints, iris, load optimization from fine-grained Smart Metering measurements, tele-diagnosis and analysis of medical signals like Electrocardiograms or DNA, to name just a few. Many of the most prominent signal processing applications deal with very sensitive signals which must not be leaked to unauthorized parties. Thus, with the advent and widespread use of outsourced computation paradigms like Cloud computing services, the challenge of protecting the signals while they are processed becomes much harder.

The field of Secure Signal Processing (SSP), also known as Signal Processing in the Encrypted Domain (SPED), was born to address these challenges, by devising efficient solutions stemming from the collaborative efforts of cryptography and signal processing. These solutions involve the use of Homomorphic Encryption (HE) as a basic building block to enable encrypted processing, but they are usually limited to additive homomorphisms like Paillier~\cite{Pa99}, and hence they also need interactive protocols in which the client (or an authority in which the client delegates trust) must communicate with the outsourced processing party in order to produce a result~\cite{TP11}. This imposes many restrictions on the client side, and presents an insurmountable barrier to the development of secure outsourced services. Hence, the goal of unattended secure signal processing, where the client only has to pre-process the inputs and post-process the outputs, is still an open problem.

This work addresses the aforementioned problem by providing a whole set of strategies and approaches to efficiently deal with composable unattended encrypted processing of sensitive signals, by relying on novel uses of Number Theoretic Transforms (NTTs) and appropriate pre- and post-processing techniques which enable efficient outsourced encrypted processing. Our proposal achieves a two-fold objective: replacing typical real or complex transforms for speeding up the underlying polynomial operations, and enabling an encrypted implementation of transformed processing in a flexible and efficient way. To the best of our knowledge, this is the first work that takes advantage of the polynomial structure of signals to represent them in a cryptosystem finite ring, where lattice cryptography can be very efficient, such that somewhat homomorphic cryptosystems can be leveraged to implement low-complexity and low-expansion ciphers and encrypted operations.

\subsection{Main Contributions}
\label{sec:mainmoticontr}

Before delving into the description of our proposed techniques, we briefly enumerate our contributions here in order to clarify the targets and scope of this work:

\begin{itemize}
\item We propose the use of NTTs with Proth prime numbers as an efficient way for performing ciphertext multiplications.

\item We present an efficient pre- and post-processing stage applied to the signals that allows us to perform: a) very efficient generalized convolutions with only one ciphertext multiplication, including cyclic convolutions, b) homomorphic NTTs with only one ciphertext multiplication, extensible to other typical fast transforms (like the Discrete Fourier Transform, DFT), and c) any type of generalized linear convolution together with an NTT or INTT (respectively DFT or IDFT).

\item We leverage the use of the relinearization primitive as a means to perform the pre- and post-processing homomorphically. Hence, we reduce the intervention of the secret key owner in the middle of the process, allowing for a set of unattended encrypted signal processing applications. We also present several optimizations to further reduce key-owner intervention: a) embed both the pre- and post-processing inside the homomorphic calculation, b) enable component-wise multiplications together with encrypted linear convolutions without an intermediate decryption, c) improve the efficiency and cipher expansion through batching/unbatching procedures.

\item We introduce and discuss a set of exemplifying encrypted signal processing applications which can be performed thanks to our novel mechanisms, comprising, among others: elementary signal processing operations (shifts, changes in sampling rate, reflections, modulations), matrix multiplications, Cyclic Redundancy Check (CRC) codes, linear transforms and interleaving operations.
\end{itemize}

\subsection{Notation and structure}
\label{sec:notation}

We represent vectors and matrices by boldface lowercase and uppercase letters, respectively. Polynomials are denoted with regular lowercase letters, ignoring the polynomial variable (e.g., $a$ instead of $a(x)$) whenever there is no ambiguity. When needed, we also represent polynomials as column vectors  of their coefficients $\bm{a}$;
$\bm{a} \cdot \bm{s}$ represents the scalar product between the vectors $\bm{a}$ and $\bm{s}$, whose components may belong to the integers or to a polynomial ring.
$R_q[x] = \mathbb{Z}_q[x]/(f(x))$ denotes the polynomial ring in the variable $x$ modulo $f(x)$ with coefficients belonging to $\mathbb{Z}_q$.

The rest of the article is structured as follows: Section~\ref{sec:prelims} briefly reviews some preliminary notions and basic cryptographic concepts needed to develop the proposed approaches. Section~\ref{sec:NTTsSSP} introduces the use of NTTs together with an optimal choice of parameters for enabling secure signal processing applications; Section~\ref{sec:generalizeETs} presents an approach to generalize convolutions and filtering in the encrypted domain; Section~\ref{sec:optimizations} proposes a series of optimizations to increase the efficiency of typical outsourced operations, and Section~\ref{sec:applications} exemplifies the use of the proposed techniques and primitives to produce a wide range of essential composable building blocks for unattended secure signal processing.

\section{Preliminaries}
\label{sec:prelims}
The majority of the traditional SSP approaches make use of additive cryptosystems like Paillier~\cite{Pa99}, which enables the calculation of additions between encrypted values by multiplying their encryptions; however, additive homomorphisms lack flexibility for tackling more complex and non-linear operations. Hence, the use of lattice cryptosystems which present a ring homomorphism (addition and multiplication) is being progressively adopted by researchers in the field \cite{TGP13, ABFK16, DGLLNW16, PTP16}; an example is Lauter's cryptosystem~\cite{LNV11}, a Somewhat Homomorphic Encryption (SHE) based on the Ring Learning with Errors (RLWE) problem that can evaluate a bounded number of consecutive encrypted operations. Other recent representative RLWE-based examples are FV \cite{FV12} and YASHE \cite{BLLN13}, cryptosystems that outperform Lauter's in terms of both efficiency and the upper bound on the number of encrypted operations. Moreover, novel lattice cryptosystems advance further in the direction of efficient processing and multi-key operation~\cite{LTV13}, and also in the fast execution of \emph{bootstrapping} for achieving true Fully Homomorphic Encryption~\cite{DM14} (FHE).

We revisit here an RLWE lattice-based cryptosystem, which we use for our mechanisms, discussing the security properties of lattice cryptosystems and the choice of parameters; we also revise the basic form of Number Theoretic Transforms (NTTs), which are the building blocks that we use to produce efficient secure signal processing primitives.

\subsection{RLWE-based cryptosystem}
For the sake of the exposition, we have chosen Lauter~\cite{LNV11} to showcase our proposed mechanisms, but they can be easily applied to any other RLWE-based cryptosystem (a brief comparison of some of the most recent homomorphic cryptosytems together with some additional reasons for choosing the Lauter cryptosystem can be found in Section~\ref{comparative-crypto}). For completeness, a slightly adapted definition of the RLWE problem particularized to the case of Lauter cryptosystem is presented:\begin{defin}[RLWE problem~\cite{LPR13}]
Given a polynomial ring $R_q[z] = \mathbb{Z}_q[z] / (1 + z^n)$ and an error distribution $\chi[z]\in R_q[z]$ that generates small-norm random polynomials in $R_q[z]$, RLWE relies upon the computational indistinguishability between samples $(a_i, b_i = a_i s + t\cdot e_i)$ and $(a_i , u_i)$, where $a_i$, $u_i$ $\leftarrow R_q[z]$ are chosen uniformly at random from the ring $R_q[z]$, while $s,e_i \leftarrow \chi[z]$ are drawn from the error distribution, and $t$ is relatively prime to $q$.
\end{defin}

The fundamental primitives and parameters of Lauter's cryptosystem are described in Table~\ref{tab:lauter_description}. Lauter's ciphertexts are composed of at least $2$ polynomial elements belonging to the ring $R_q[z]$; the cryptosystem allows for additions (the smallest ciphertext is previously zero-padded) and multiplications on these tuples of polynomials, whose size is increased after each multiplication (the original size can be brought back by resorting to the relinearization operation, explained in Section~\ref{sec:relinearization}). 
The security of the cryptosystem is based on the hardness of reducing the $n$-dimensional lattices generated by the secret key and also on the semantic security provided by the RLWE problem (two encryptions of the same or different messages are indistinguishable). Further details about possible attacks to the cryptosystem are included in Section~\ref{sec:security}.

With this cryptosystem, messages encoded as univariate polynomials can be encrypted in only one ciphertext (instead of encrypting each coefficient in a different ciphertext). This has the main advantage of enabling to homomorphically perform encrypted linear convolution operations in a natural way with only one multiplication between ciphertexts; there is only a small overhead due to the larger cardinality of the involved encrypted polynomial coefficients, which belong to $\mathbb{Z}_q$ instead of the plaintext $\mathbb{Z}_t$, with $q>t$. In order to allow for $D$ consecutive products and $A$ sums over the same ciphertext, the needed $q$ for correct decryption is lower-bounded by
\begin{equation}
q \ge 4(2t \sigma^{2} \sqrt{n})^{D+1}(2n)^{D/2}\sqrt{A}.
\label{eq:condq}
\end{equation}
Remarkably, Lauter can be securely adapted to work efficiently with multidimensional signals (2D and 3D images or video), by extending the RLWE problem to a multi-variate case~\cite{PTP15}; this extension enables working with complex-coefficient polynomials, by using bi-variate encryptions in which one of the modular polynomials is $f(w) = 1 + w^2$. We will revisit this idea for some of our constructions.

\begin{table}[!t] 
\renewcommand{\arraystretch}{1.3}
\caption{RLWE-based Lauter Cryptosystem: Parameters and Primitives}
\label{tab:lauter_description}
\centering \scriptsize

\begin{tabular}{|m{0.4in}|m{0.3in}|m{2.3in}|}

\hline
\multicolumn{3}{|c|}{Parameters}\\
\hline
\multicolumn{3}{|p{3.3in}|}{ 
Let $R_t[z] = \mathbb{Z}_t[z] / (1 + z^n)$ be the cleartext ring and $R_q[z] = \mathbb{Z}_q[z] / (1 + z^n)$ the ciphertext's. The noise distribution $\chi[z]$ in $R_q[z]$ takes its coefficients from a spherically-symmetric truncated i.i.d Gaussian $\mathcal{N}(\bm{0},\sigma^2\bm{I})$; $q$ is a prime $q \equiv 1 \bmod{2 n}$, and $t<q$ is relatively prime to $q$.}\\\hline

\multicolumn{3}{|c|}{Cryptographic Primitives}\\

\hline
SH.KeyGen& Process & $s, e \leftarrow \chi[z]$, $a_1 \leftarrow R_q[z]$  $sk = s$ and $pk = (a_0 = -(a_1 s + te), a_1)$ \\

\hline
\multirow{2}{*}{SH.Enc}
& Input & $pk = (a_0, a_1)$ and $m \in R_t[z]$ \\
\cline{2-3}
& Process & $u, f, g \leftarrow \chi[z]$ and the fresh ciphertext is $\bm{c} = (c_0, c_1) = (a_0 u + tg + m, a_1 u + tf)$ \\

\hline
\multirow{2}{*}{SH.Dec}
& Input & $sk$ and $\bm{c} = (c_0, c_1, \dots,c_{\gamma-1})$ \\
\cline{2-3}
& Process & $m = \left( \left(\sum_{i = 0}^{\gamma-1} c_i s^i \right) \bmod q \right) \bmod{t}$ \\

\hline
\multirow{2}{*}{SH.Add}
& Input & $\bm{c}_0 = (c_0, \dots, c_{\beta-1})$ and $\bm{c}_1 = (c_0', \dots, c_{\gamma-1}')$ \\
\cline{2-3}
& Process & $\bm{c}_{add} = (c_0 + c_0', \dots, c_{\max{(\beta, \gamma)}-1} + c_{\max{(\beta, \gamma)}-1}')$  \\

\hline
\multirow{2}{*}{SH.Mult}
& Input & $\bm{c}_0 = (c_0, \dots, c_{\beta-1})$ and $\bm{c}_1 = (c_0', \dots, c_{\gamma-1}')$ \\
\cline{2-3}
& Process & Using a symbolic variable $v$ their product is $\left( \sum_{i=0}^{\beta-1} c_i v^i \right) \cdot \left( \sum_{i=0}^{\gamma-1} c_i' v^i \right) = \sum_{i=0}^{\beta + \gamma-2} c_i'' v^i$ \\

\hline
\end{tabular}
\vspace{-0.2cm}
\end{table}

\subsubsection{Choice of the RLWE-based cryptosystem}
\label{comparative-crypto}

Although we have chosen the Lauter cryptosytem as the basis for our proposals, any RLWE-based cryptosytem can be used in order to apply the proposed methodologies and tools. The only requirement is the use of a modular function of the form $f(z) = 1 + z^n$ which, in fact, seems to be the most accepted and widely used by the cryptographic community due to its efficiency and well studied properties.

Besides its simplicity, there are some interesting motivations for our choice of the Lauter cryptosystem. Costache and Smart~\cite{CS16} recently presented a comparison in terms of efficiency, cipher expansion and security of the four main variants of RLWE-based cryptosytems: the NTRU and BGV schemes, which encode the messages in the lower bits of the decryption equation, and their corresponding scale-invariant versions YASHE and FV, encoding the messages in the upper bits of the decryption equation. They show that the most efficient schemes for the case of small and large moduli in the plaintext coefficients are respectively YASHE and BGV cryptosystems, where the former performs only slightly better than BGV for very small plaintext moduli ($t=2$).
Therefore, we chose Lauter cryptosystem as a representative of the BGV family, as a large number of signal processing applications work with reasonably large signal values and some of our contributions assume a relatively large value for $t$.

Finally, a recent attack \cite{ABD16} against NTRU cryptosytems also affects YASHE for some practical values that were considered secure until now. Considering $\lambda$ as the security parameter, this attack allows to break these cryptosytems in sub-exponential time in $\lambda$ for super-polynomial $q(\lambda)$; and even in polynomial time when $q(\lambda)$ increases. As this attack has no known effect on the BGV cryptosystem, our choice seems to be the most suitable thanks to both its efficiency and security.

\subsection{Security of Lattice-based Cryptosystems}
\label{sec:security}

This section revisits some practical aspects related to security of lattice cryptosytems. The underlying assumption supporting the security of the used cryptosystems is the indistinguishability of the RLWE distribution w.r.t. a uniform distribution. There are mainly two types of attacks that can be considered: a) distinguishing attacks~\cite{MR09}, whose goal is to break the indistinguishability assumption through basis reduction algorithms, and b) decoding attacks, which are aimed at obtaining the secret key $s$. We focus on the former.

Although we do not specifically account for decoding attacks, by using values for $n$ similar to those used in~\cite{LNV11}, the cryptosystem achieves protection against them as described in \cite{LP11}. Therefore, we adhere to these minimum values for $n$.

\subsubsection{Security and runtime attack as a function of the root Hermite factor $\delta$}
\label{sec:rootHermite}
The best attacks against lattice cryptosystems rely on basis reduction algorithms. Given an arbitrary basis of a lattice, these algorithms try to obtain a nearly orthogonal basis with small vectors. Among them, BKZ~\cite{BKZ2} is currently one of the most efficient ones. It uses blocks of size ranging from $2$ to the dimension of the lattice; increasing block sizes produce better bases at the cost of a higher computational load.

We take as a commonly adopted measure of security the root Hermite factor $\delta > 1$ for the underlying lattice, which is directly related to the running time needed for a basis-reduction algorithm to succeed. In fact, the runtime of an attack is approximately proportional to $e^{k/\log\delta}$ for a constant $k$; i.e., a lower $\delta$ implies a higher security. For the optimal distinguishing attack using BKZ, we obtain the following expression for $\delta$~\cite{LNV11}:
\begin{equation*}
\label{eq:delta}
\log_{2}(\delta) = {(\log_{2}(c \cdot q/s))}^{2}/(4 n \log_{2}(q)),
\end{equation*}
where $n$ is the rank of the lattice, $c \approx \sqrt {\ln (\frac{1}{\epsilon})/\pi}$, $\epsilon$ is the attacker advantage ($\epsilon=2^{-32}$), and $s$ is a scale parameter of the error distribution (for the $n$-dimensional Gaussian $s = \sigma \sqrt{2 \pi}$).

In order to calculate the corresponding bit security, we resort to the lower bound estimate $t_{BKZ}(\delta)$ of \cite{LP11}:
\begin{equation}
t_{BKZ}(\delta) = \log_2{\left( T_{BKZ}(\delta) \right)} = \frac{1.8}{\log_2{\delta}} - 110.\label{eq:bit-sec-rlwe}
\end{equation}

For the other cryptosystem parameters, we choose $\epsilon = 2^{-32}$, $\sigma = 1$ and for $q$, we choose the smallest prime that satisfies the bound~\eqref{eq:condq}, where we have $A = 1$, $D = 1$ and $t = 65537$.
Table~\ref{tab:lautercrypto} shows different runtimes and the relevant security parameters ($\delta$ and bit security) of the used cryptosystem.

\begin{table}[!t]
\renewcommand{\arraystretch}{1.3}
\caption{Performance of Lauter cryptosystem ($D=1$, $A=1$, $t=65537$, $s=\sqrt{2\pi}$) and Paillier cryptosystem}
\label{tab:lautercrypto}
\label{tab:pailliercrypto}
\centering \scriptsize
\begin{tabularx}{\columnwidth}{|>{\setlength\hsize{1\hsize}\centering}X| c c c c c|}\hline
\multicolumn{6}{|c|}{Lauter cryptosystem}\\\hline
$n$ & $1024$ & $2048$ & $4096$ & $8192$ & $16384$ \\
\hline 
\hline
 $\lceil \log_2{(q)} \rceil$ & $53$ & $55$ & $56$ & $58$ & $59$ \\
 $\delta$ & $1.0090$  & $1.0046$  & $1.0024$  & $1.0012$  & $1.0006$ \\
 Bit security (Eq.\eqref{eq:bit-sec-rlwe}) & $\approx 30$  & $\approx 162$  & $\approx 410$  & $\approx 930$  & $\approx 1270$ \\
 Encrypt. time ($\mu$\textit{s}) & $114$ & $224$ & $444$ & $860$ & $1780$ \\
 Decrypt. time ($\mu$\textit{s}) & $37$ & $77$ & $159$ & $326$ & $695$ \\
 Poly. Mult. time ($\mu$\textit{s}) & $24$ & $46$ & $91$ & $171$ & $353$ \\
 Poly. Add. time ($\mu$\textit{s}) & $7$ & $12$ & $24$ & $47$ & $107$ \\
 Pre/Post time ($\mu$\textit{s}) & $4$ & $8$ & $17$ & $32$ & $64$ \\
\hline\hline
\end{tabularx}

\begin{tabularx}{\columnwidth}{|>{\setlength\hsize{1\hsize}\centering}X| c c c c|}\hline
\multicolumn{5}{|c|}{Paillier cryptosystem}\\\hline
Modulus size (bits) & $1024$ & $2048$ & $3072$ & $7680$ \\
\hline 
\hline
 Bit security & $\leq 80$  & $112$  & $128$  & $192$ \\
 Encrypt. time ($\mu$\textit{s}) & $2947$ & $19438$ & $54122$ & $521981$  \\
 Decrypt. time ($\mu$\textit{s}) & $2806$ & $19269$ & $54006$ & $521761$  \\
 Scalar Mult. time ($\mu$\textit{s}) & $27$ & $92$ & $182$ & $729$  \\
 Scalar Add. time ($\mu$\textit{s}) & $3$ & $10$ & $21$ & $87$  \\
\hline
\end{tabularx}
\vspace{-0.2cm}
\end{table}

\paragraph{Paillier cryptosystem performance}
Paillier cryptosystem~\cite{Pa99} has been extensively used in recent years for secure signal processing. Therefore, we compare the efficiency of our proposed solutions exploiting Lauter with typical solutions resorting to Paillier. Due to the different hardness problems in which both cryptosystems are based on, we base our fair comparisons on the bit-security that can be achieved with both schemes. Table~\ref{tab:pailliercrypto} reports the corresponding runtimes and bit security for different modulus size of the Paillier cryptosystem  (with plaintext values upper-bounded by $256$). For Paillier, we resort to the bit security estimate of RSA~\cite{NISTreport}. It is also important to note that Paillier can only deal with one scalar plaintext, while all the primitives using Lauter work in parallel with $n$ plaintext values encrypted in one ciphertext, each one encoded in a different coefficient of the polynomials in $R_t[z]$, which is a clear advantage.
Despite that, Table~\ref{tab:pailliercrypto} shows that the encryption of one scalar with Paillier is much slower than the time needed to perform our proposed pre-/post-processing (see Section~\ref{sec:generalizeETs}) and the encryption of $n$ numbers with Lauter.

\subsection{Number Theoretic Transforms (NTTs)}
\label{sec:NTTs}
Signal processing heavily relies on transformed processing, for which the usually employed transforms are based on the DFT (Discrete Fourier Transform), due to the physical meaning of the frequency domain, the efficient algorithms for their computation, the good energy compaction properties, and the possibility of taking advantage of the cyclic convolution property. The latter implies a correspondence between the cyclic convolution of two signals and the element-wise product of their transforms, enabling very efficient computation of convolutions by working in the transformed domain.

When dealing with secure encrypted processing, DFTs cannot be directly translated to the encrypted signals, due to their reliance on complex arithmetic and non-integer numbers, which cannot fit in the finite rings of the cryptosystems without a quantization; this poses subsequent problems of accuracy loss and scale factor accumulation (cipher blow-up).
When working in finite rings, we can find an alternative approach by resorting to integer transforms, more amenable to encrypted processing: NTTs (Number Theoretic Transforms) are transforms with the same structure as the DFT, with the peculiarity that they operate with elements belonging to a finite field or ring instead of the complex field. 

More formally, in a finite ring $R_p = Z_p[z] / f (z)$ with $p = \prod_{i=1}^K p_i^{m_i}$, an NTT of size $N$ can be defined (with the cyclic convolution property) if the following properties hold \cite{Nussbaumer82}:

\begin{itemize}
\item There exists an $N$-th root of unity $\alpha$ in $R_p$, for which $\gcd(\alpha, p) = \gcd(N, p) = 1$.

\item $N$ is a divisor of $\gcd \left( p_1 - 1, \ldots, p_K - 1\right)$.
\end{itemize} 

The expressions for the forward and inverse transforms are
\begin{align} 
X[k] =& \sum_{l = 0}^{N-1} x[l] \alpha^{lk} \bmod{p},\; k = 0, 1, \dots, N - 1\label{eq:ntt}\\
\nonumber
x[l] =& N ^{-1} \sum_{k = 0}^{N-1} X[k] \alpha^{-lk} \bmod{p}, \;n = 0 , 1, \dots, N - 1.
\end{align}

Analogously to DFTs, NTTs possess a cyclic convolution property, and they also enable fast computation algorithms like radix-$2$ and radix-$4$. Remarkably, the NTTs lived a golden age in Signal Processing when the available hardware at the time (FPGAs and DSPs) could only work with finite precision arithmetic, but were later replaced due to the generalization of floating-point-capable hardware.

For our purposes, their important property is that they work in the same integer rings as lattice cryptosystems do and, therefore, they impose no rounding errors or cipher blow-up. 
Consequently, NTTs can be used to efficiently perform polynomial multiplications, and they have been recently proposed as a means to speed up finite-ring polynomial multiplications: there are some cryptosystem realizations that make use of NTTs for improving the efficiency of their polynomial operations \cite{APS13,PG14,RVMCV14,CMVRCPV15,DHS15}.
Our proposed techniques go further, by focusing on an unexplored specific subset of the available NTTs and adjusting the cryptosystem parameters accordingly, to produce new primitives that enable highly efficient implementations, as we show in the next sections.

\section{Number Theoretic Transforms in Secure Signal Processing}
\label{sec:NTTsSSP}
Once we have introduced the notions of lattice-based somewhat homomorphic cryptosystems and the basic formulation for NTTs, we discuss the proposed setup for the optimal combination of these two concepts with the goal of achieving extremely efficient, unattended outsourced signal processing.

We particularize the NTT for its application to the cryptosystems presented in \cite{LNV11} and \cite{PTP15}. In that case, the cryptosystems require the use of a ring $R_q = Z_q[z] / (1 + z^n)$, where $q$ is prime, and $n$ is a power of $2$. Additionally, $q$ must verify $q \equiv 1 \mod{2n}$ and meet the lower bound defined by the number of operations allowed on the same ciphertext, cf. Eq.~\eqref{eq:condq}. 
Therefore, combining these restrictions with the existence of the size-$n$ NTT in the ring $R_q$, we have 
\begin{itemize} 
\item $n$ divides $\phi(q)$, where $\phi(q) = q - 1$. 
\item $q$ verifies both Eq.~\eqref{eq:condq} and $q \equiv 1 \mod{2n}$. 
\end{itemize} 
Prior works simply assume that such a prime exists and do not address its generation or adaptation to efficient processing with a given cryptosystem. We can prove that these restrictions are verified by the set of Proth primes \cite{Baillie79,Robinson58,RTKH77}, which can be easily generated. A Proth number $q$ is characterized by the form $q = k 2^l + 1$ where $l$ is an integer, $k$ is an odd positive integer and $2^l > k$.
The primality test for Proth numbers follows by virtue of Proth's theorem:
\noindent
\begin{thProth}[\cite{Proth1878}]
For a Proth number $q$, if there is at least an integer $a$ satisfying $a^{(q-1)/2} \equiv -1 \mod{q}$, then $q$ is prime. 
\end{thProth}

Once $q$ is fixed, an $n$-th root of unity $\alpha$ can be found by searching those numbers that have an order greater than or equal to $n$ in the set of integers $[1, q-1]$. If the found $\alpha$ has an order $d$ higher than $n$, then the $n$-th root of unity is obtained by considering $\alpha^{(\frac{d}{n})n} \equiv 1 \mod{(q)}$, where $\alpha^{\frac{d}{n}}$ is an $n$-th root of unity. Algorithm~\ref{pseudocodeProth} details this procedure.

\begin{algorithm}
  \caption{Cryptosystem's Parameters ($q$, $n$, $\alpha$) Generation }
  \label{pseudocodeProth}
\begin{algorithmic}[1]
\Procedure{Search Cryptosystem Parameters}{N}
\\
\emph{Input}:
\State $N \gets \text{desired size for the NTT}$ \textit{ (where N is a power
\State  of 2)}
\\
\emph{Search prime $p$ for the NTT of size $N$}:
\While{$p$ not found}
\State Find $p$ of the form $p = k2^l + 1$ \textit{(with $k$ and $l$ natural numbers, where $k$ is not a power of $2$)}
\If {$p$ is prime and $N$ divides $p-1$}
\State \textbf{break};
\EndIf
\EndWhile
\\\emph{Search $N$-th root of unity $\alpha$}:
\For{$i = 1$ until $p-1$}
\If{$i$ is a $M$-th root of unity with $M \geq N$}
\State $\alpha \gets i^{M/N}$
\State \textbf{break};
\EndIf
\EndFor
\\
\emph{Output}:
\State For the ring $R_q$ we have $q = p$ and $n = N$ 
\State An $N$-th root of unity $\alpha$
\EndProcedure
\end{algorithmic}
\end{algorithm}

This choice of parameters enables efficient cyclic convolutions between the ciphertext elements with no rounding errors, as they allow for efficient algorithms for NTT and INTT (e.g., radix-$2$ or radix-$4$). The particularity here is that the cyclic convolutions allowed by this setting are actually nega-cyclic, and further processing has to be applied to enable ``regular'' cyclic convolutions, as we explain in the next section.

\section{Generalizing Convolutions and Transforms in the Encrypted Domain}
\label{sec:generalizeETs}
Equipped with the presented description of RLWE cryptosystems and the proposed optimal parameters for NTTs, we detail here the main contribution of this work, comprising a versatile set of novel secure signal processing primitives:

\begin{itemize}
\item We show how to efficiently perform any encrypted cyclic, negacyclic or generalized convolution in an RLWE-based cryptosystem in a more efficient way and without wasting any coefficient. For that purpose, we propose an efficient pre- and post-processing for the input signals and the result respectively, enabling further operations in the encrypted domain.
\item In order to allow for transformed operations under encryption, we propose a practical method for computing an encrypted NTT or DFT with an RLWE-based cryptosystem.
\item Finally, both results are generalized in a framework that enables any kind of encrypted convolution and linear transforms with a convolution property.
\end{itemize}

\subsection{Encrypted Cyclic Convolution} 
\label{cycliclauter}

Along with the linear convolution, the circular or cyclic convolution is frequently used in signal processing. To implement a linear convolution with an RLWE cryptosystem \cite{LNV11, BV11b} and its extensions \cite{PTP15}, we need a value of $n$ large enough to store the result of the convolution. Moreover, the cyclic convolution poses additional problems as the modular function $f(z) = 1 + z^n$ in the cryptosystem only allows for negacyclic convolutions~\cite{Dav12}. A straightforward approach for calculating the cyclic convolution would be the following: 

\begin{itemize}
\item The cryptosystem modular function is of the form $f(z) = 1 + z^N$, with $N$ power of two.
\item The larger signal is assumed to have length $N/2$ (in other case, it is zero padded).
\item By the homomorphic properties of the cryptosystem, the allowed polynomial multiplication enables computing $x(z)h(z) (1 + z^{N/2}) \bmod{(1 + z^N)}$.
\end{itemize} 

It can be shown that the output of this product holds the result of the negacyclic convolution in the $N/2$ first coefficients, and the result of the cyclic convolution in the last $N/2$ coefficients. The drawbacks are that: a) half of the used coefficients are wasted, unnecessarily increasing cipher expansion, and b) the result is located in a portion of the ciphertext, so reusing it for further operations becomes harder.

We present our method for calculating the encrypted cyclic convolution, by using just one polynomial product and element-wise pre- and post-processing. This approach yields a more efficient cipher expansion, and it also enables to continue performing operations with the results of the convolutions.

\subsubsection{Efficient Pre- and Post-processing}
\label{sec:prepostproc}
We rely on a generalization of the cyclic convolution between two signals $x[l],h[l]$ in terms of a complex value $\alpha$, proposed by Murakami~\cite{Murakami00}. The $\alpha$-generalized cyclic convolution is defined as
\begin{equation} 
y(z) = x(z)h(z) \bmod{(1 - \alpha z^n)},\label{eq:generconv}
\end{equation}
where $x(z)$, $h(z)$ and $y(z)$ are the $Z$-transforms of $x[l]$, $h[l]$ and $y[l]$ (we use the definition for the $Z$-transform as a power series in $z$ instead of the more common $z^{-1}$).

This generalization lets us specify different types of convolutions depending on the chosen $\alpha$: for $\alpha = -1$ we obtain a negacyclic convolution (we refer the reader to \cite{Dav12} for more details on negacyclic convolutions), which corresponds to the homomorphic operation offered by RLWE-based cryptosystems with $f(z) = 1 + z^n$. Conversely, $\alpha = 1$ conforms to the cyclic or circular convolution. We aim at a regular cyclic convolution ($\alpha=1$), but we are bound to a negacyclic one by the cryptosystem homomorphism, as a modular function of the form $f(z)=1-z^n$ would not be irreducible in the integers (see \cite{MR09, SS11} for further details on the security reasons behind discarding $1 - z^n$ as the modular function).

Supported by Murakami's formulation, we can enable the calculation of a cyclic convolution between two $N$-length polynomials $x[l]$ and $h[l]$ by carrying out the following steps: 

\begin{itemize} 
\item Prior to encryption, the input signals are pre-processed with component-wise products: 
\begin{align*} 
x'[l] =& x[l] \alpha^{-l/N} {(-1)}^{l/N} Q,\;l=0,\ldots,N-1, \\ 
h'[l] =& h[l] \alpha^{-l/N} {(-1)}^{l/N} Q,\;l=0,\ldots,N-1, 
\end{align*} 
where $Q$ is the quantization applied to signals $x'[l],h'[l]$. 
\item Then $y'(z)$ can be calculated under encryption with a homomorphic polynomial product
\begin{equation*} 
y'(z) = x'(z) h'(z) \bmod{(1 + z^N)} 
\end{equation*} 
\item The output decrypted signal $y'[n]$ is post-processed 
\begin{equation*} 
y[l] = y'[l] \frac{\alpha^{l/N} {(-1)}^{-l/N}}{Q^2}.
\end{equation*} 
\end{itemize} 

With the described procedure, the $\alpha$-generalized cyclic convolution can be successfully implemented with a single product of encrypted polynomials; in particular, the cyclic convolution can be implemented with $\alpha = 1$.

It must be noted that the element-wise product between the pre- and post-processing vectors is equal to a constant signal of ones, so the polynomial product between two pre-processed polynomials ``preserves'' the pre-processing and can be subsequently operated. Consequently, we can implement several products between ciphertexts and the results of ciphertext products without the intervention of the key owner in the middle of the process. That is, for performing several cyclic convolutions we only need the key owner to apply the element-wise pre-processing to the original encrypted signals.

A final remark must be made regarding the complex arithmetic assumed by Murakami's formulation. The $N$ roots of $-1$ needed for building the pre- and post-processing vectors can be tackled in two ways: a) complex numbers can be embedded in the cryptosystem by incorporating a modular function $f(w) = 1 + w^2$ to the multivariate ring~\cite{PTP15}, and b) Murakami's concepts can be applied to finite rings, so that $\alpha^{1/N}$ and ${(-1)}^{1/N}$ are elements of $\mathbb{Z}_t$, and the conditions for the existence of the $N$-point NTT are still satisfied.
In that case, we could discard the quantization $Q$ and perform the encrypted $\alpha$-generalized convolution without rounding errors.

Consequently, we have solved the two main limitations that current approaches have for calculating a cyclic convolution under encryption: our approach does not introduce any rounding errors, it does not need to discard any coefficient (reducing the effective cipher expansion), and it can cope with successive operations without intermediate decryptions. We evaluate now its performance in terms of computational complexity.

\paragraph{Performance evaluation of the encrypted cyclic convolution}
\label{eval1}

\begin{figure}[!tb]
  \centering
  \vspace{-0.38cm}
\subfigure[Encrypted cyclic convolution runtimes.]{\label{fig:conv1}\includegraphics[height=2.8in, width = 3.44in]{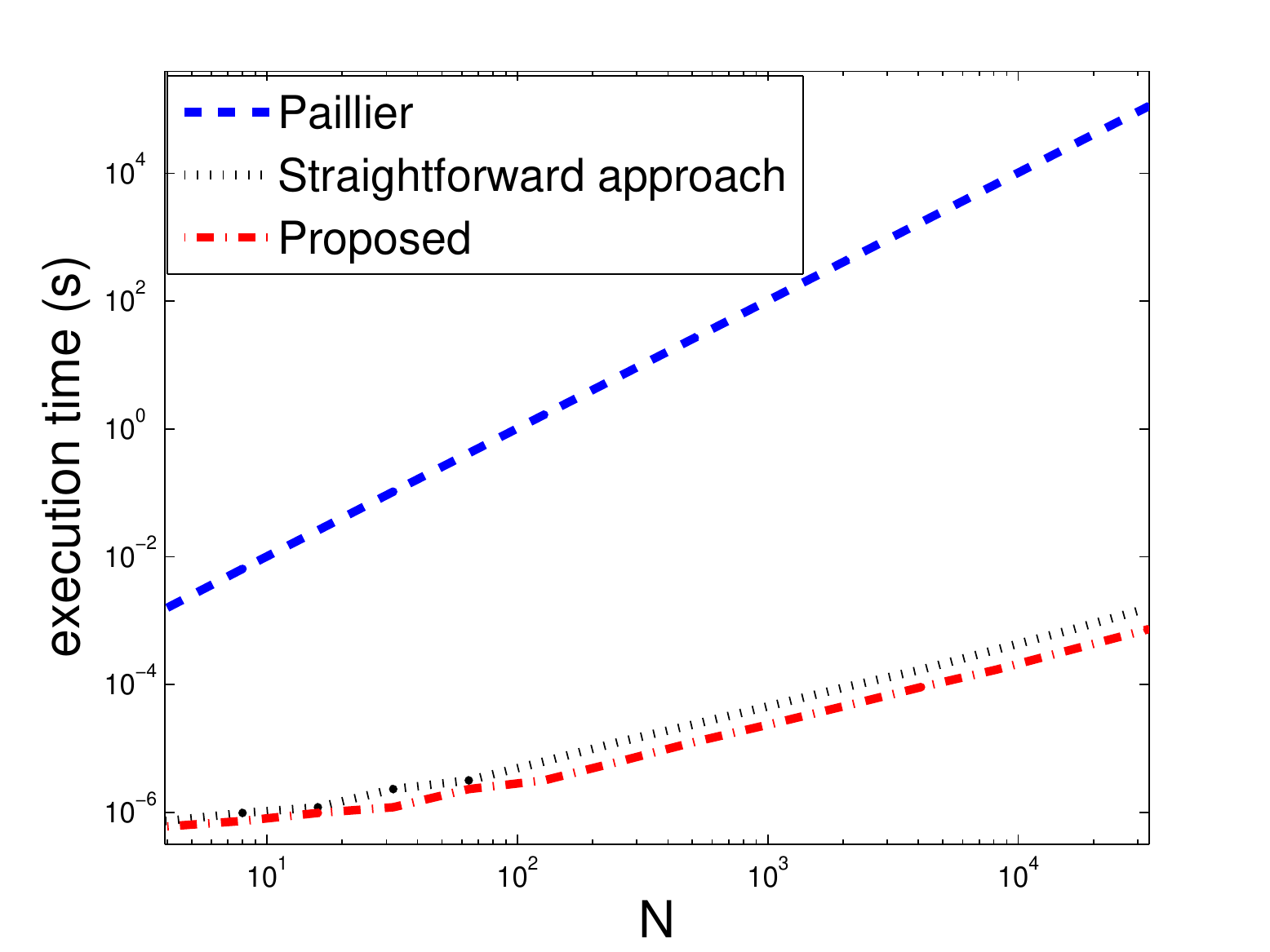}}
\subfigure[Encryption/decryption runtimes for cyclic convolution.]{\label{fig:conv2}\includegraphics[height=2.8in, width = 3.44in]{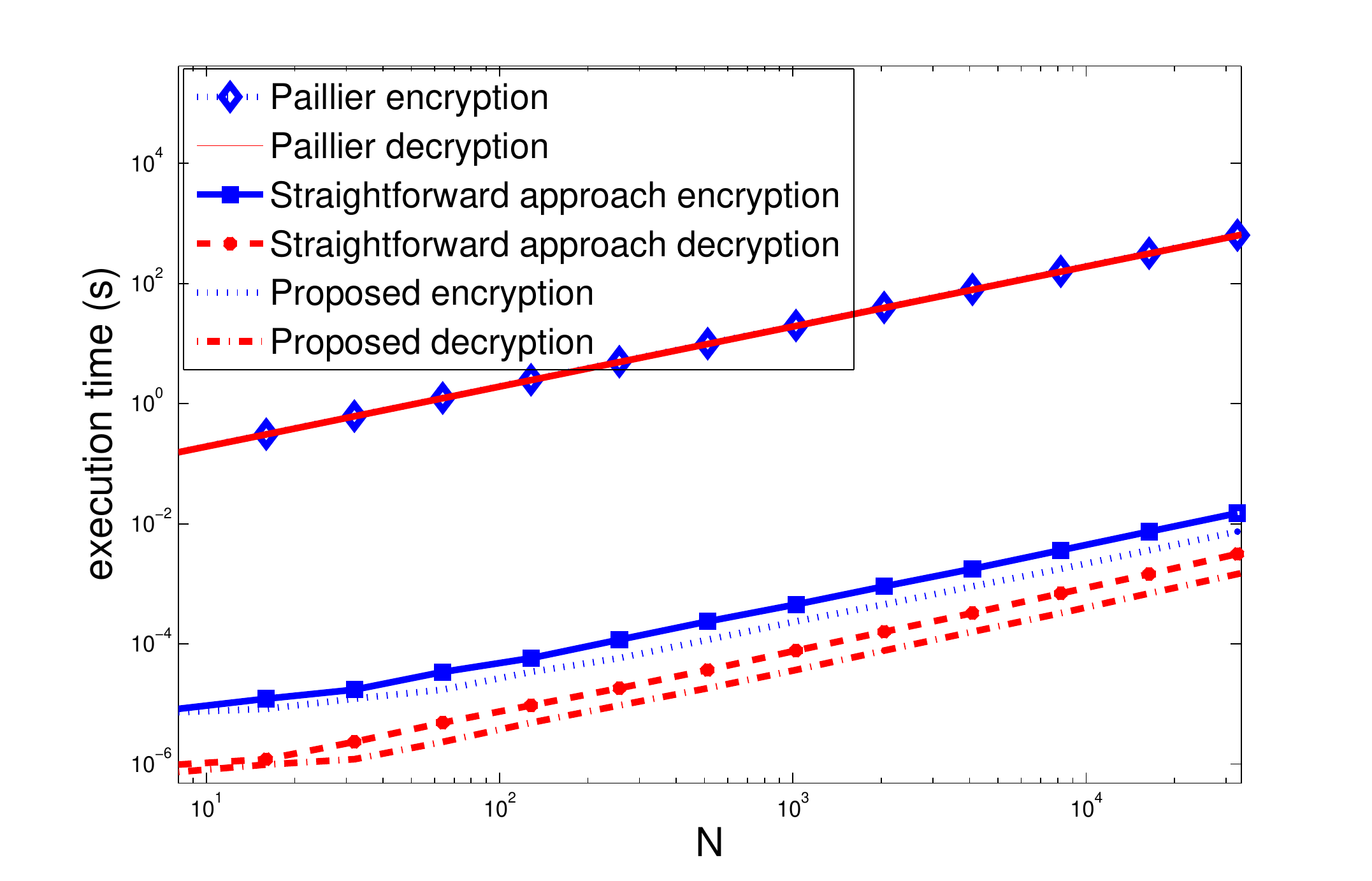}}\\
\vspace{-0.38cm}
\subfigure[Encrypted NTT runtimes.]{\label{fig:ntt1}\includegraphics[height=2.8in, width=3.44in]{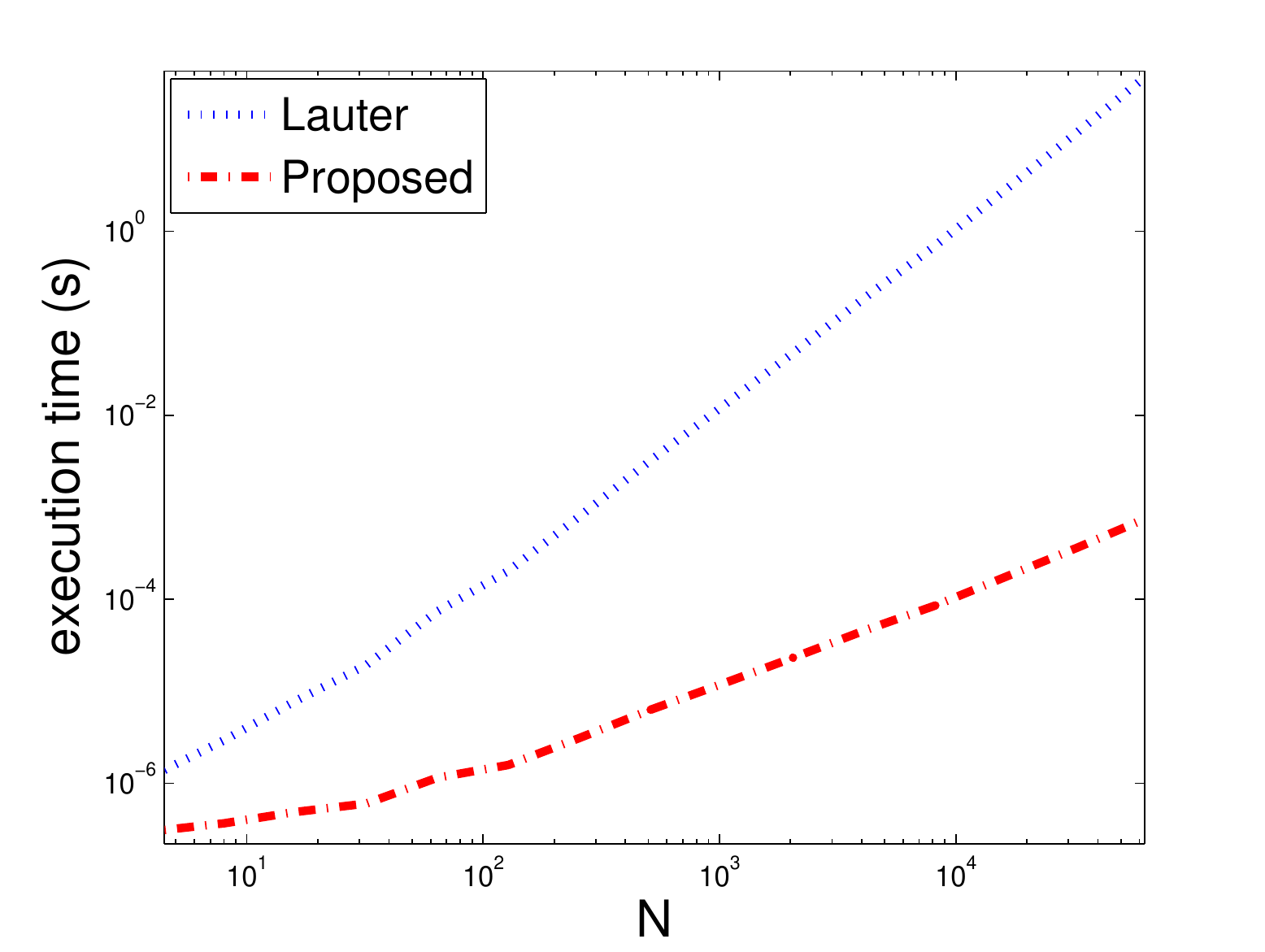}}
\subfigure[Encryption/decryption NTT runtimes.]{\label{fig:ntt2}\includegraphics[height=2.8in, width=3.44in]{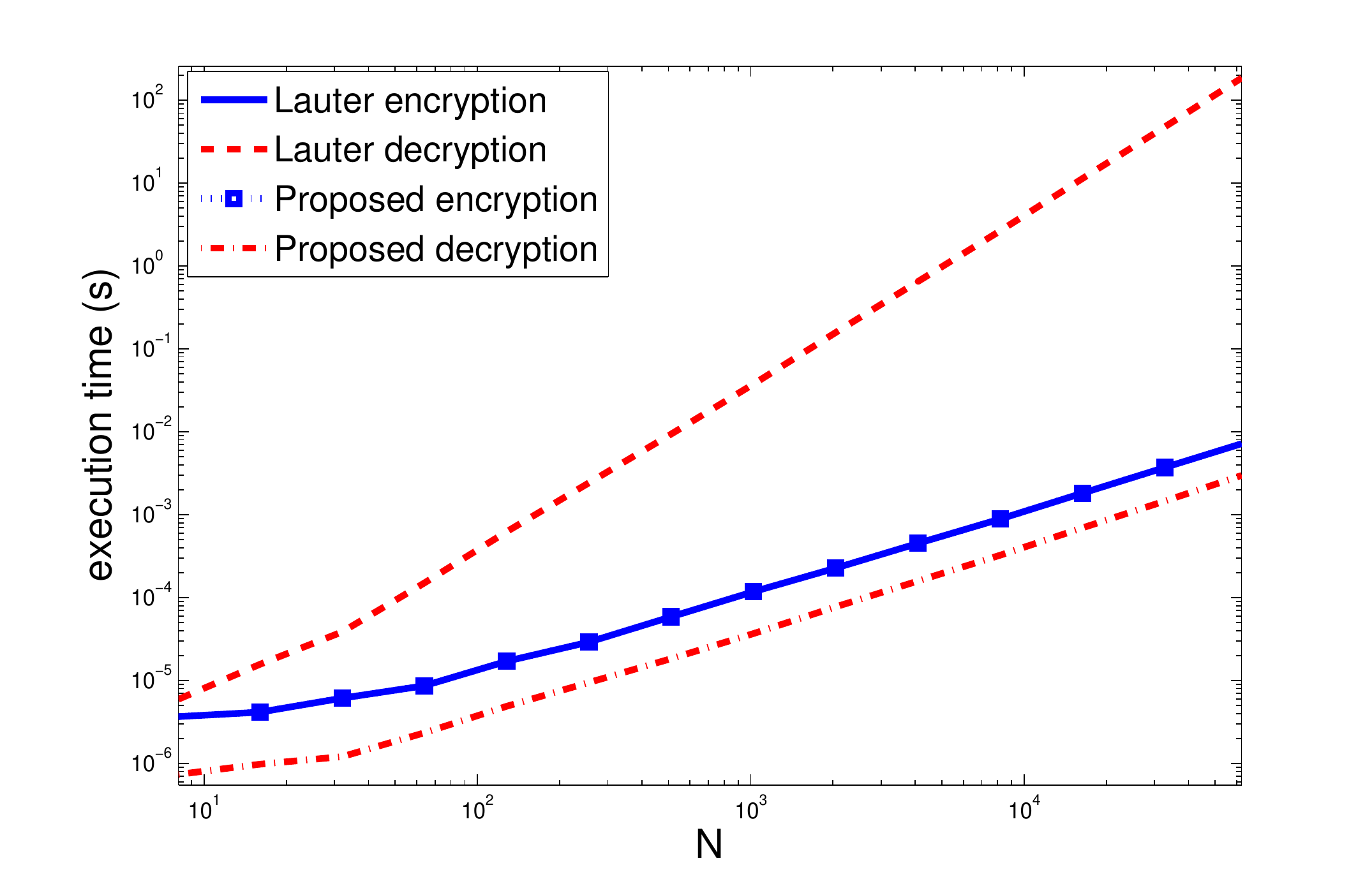}}
\vspace{-0.1cm}
\caption{Performance of encrypted cyclic convolutions and NTT.}
\label{fig:exp12}
\vspace{-0.3cm}
\end{figure}

We have implemented Lauter's RLWE-based cryptosytem in C++ using the GMP 6.0.0 \cite{gmplib} and NFLlib \cite{ABGGKL16} libraries. Figures \ref{fig:exp12}a and \ref{fig:exp12}b compare the encrypted cyclic convolution performance with a 2048-bit modulus Paillier-based convolution (one of the convolved signals cannot be encrypted) versus the proposed method with Lauter's cryptosytem with $n=N$ on an Intel Xeon E5-2620 processor running Linux. We show the comparison of a) the encrypted convolution, and b) the encryption and decryption times with our pre- and post-processing. Additionally, the computational cost of performing a cyclic convolution of two encrypted signals with our scheme is lower than the straightforward method, due to the reduction in the needed coefficients (no coefficients are discarded).

We are not considering relinearization steps after each multiplication, but account for the decryption of the extended encryptions. We can see that RLWE-based cryptosystems are more efficient than Paillier, also having a much lower cipher expansion. Moreover, our method enables chaining several consecutive encrypted cyclic convolutions in a natural way.

Finally, we can see that in our method $n = N$; hence, when we increase the size of the encrypted signals we are also reducing the root Hermite factor $\delta$, therefore increasing the runtime of a distinguishing attack (see Section~\ref{sec:security})~and achieving a much higher bit-security than Paillier (see Table~\ref{tab:lautercrypto}).

\subsection{Encrypted NTTs}
\label{sec:encryptedNTTs} 
With the relation of equivalence between the clear and encrypted convolutions, we can use the efficient radix-$2$/radix-$4$ algorithms of the NTTs for performing the negacyclic convolutions of encrypted signals as shown in Section \ref{sec:NTTsSSP}. In this case, the NTT is applied directly on the encrypted signals as a means to speed up the calculation of the polynomial product (nega-cyclic convolution), which gets reduced to a component-wise product in the transformed domain.

However, there are cases where the NTT must be applied to the clear-text signals, and we must replicate the computation of the NTT once the signals are already encrypted, in such a way that once we decrypt we get the transformed coefficients of the clear-text signal. Therefore, we aim here at the encrypted implementation of NTTs (homomorphically applied to the cleartext) independently of whether the underlying encrypted polynomial products are implemented with the aid of NTTs. Previous works have focused only on the implementation of the encrypted DCT or DFT \cite{BPB09a}, but not on NTTs. For this purpose, we propose a mechanism to obtain the NTT of a signal under encryption with only a cyclic convolution and a pre- and post-processing step. This procedure can also be applied to any other transform with a similar structure, with the difference of having to work with rounded real or complex numbers. In fact, we could separately operate with the real and imaginary parts of the signals or even embed complex numbers in the cryptosystem~\cite{PTP15} by incorporating a modular function $f(w) = 1 + w^2$. Hence, we can use the same procedure to implement the encrypted versions of the corresponding real or complex transforms, but working over complex $2N$-th roots of unity; hence, by applying a pre- and post-processing step we get to homomorphically perform a DFT with only one cyclic convolution, or one DCT with two cyclic convolutions (for the DCT we would resort to Euler's formula to represent the cosines as a combination of complex roots of unity). It must be noted that unlike the NTT, both the DCT and the DFT would need quantization prior to encryption in order to be able to represent the real numbers as integers, with the corresponding increase in cipher expansion and quantization error.

We first introduce the proposed encrypted NTT algorithm, which we later extend for computing NTTs, INTTs and generalized cyclic convolutions.

\subsubsection{Encrypted NTT with pre- and post-processing}
\label{sec:prepostNTTs}
By resorting to the formulation of Bluestein FFT algorithm (also called chirp z-transform algorithm \cite{Bluestein70,RSR69}), we can compute the NTT of a signal as a single convolution and a pre- and post-processing. 
The expression for the NTT of a signal $x[l]$ is given in Eq.~\eqref{eq:ntt}.  
We need that $\alpha^{\frac{1}{2}}$ be a $2N$-th root of unity in $\mathbb{Z}_t$ (and hence $\alpha$ is a $N$-th root of unity in $\mathbb{Z}_t$), so that we can write $kl = - \frac{{(k - l)}^2}{2} + \frac{l^2}{2} + \frac{k^2}{2}$. Hence,
\begin{equation*} 
X[k] = \alpha^{\frac{k^2}{2}} \sum_{l = 0}^{N-1} \alpha^{\frac{l^2}{2}} x[l] \alpha^{\frac{-{(k - l)}^2}{2}}.
\end{equation*} 

This shows the equivalence to a cyclic convolution followed by a component-wise product with $\alpha^{\frac{k^2}{2}}$
\begin{equation*} 
X[k] = \alpha^{\frac{k^2}{2}}\sum_{l = 0}^{N-1} x'[l] \alpha^{\frac{-{(k - l)}^2}{2}} = \alpha^{\frac{k^2}{2}} (x'[k] \circledast \alpha^{\frac{-k^2}{2}}),
\end{equation*}
where $x'[l] = \alpha^{\frac{l^2}{2}}x[l]$ and $\circledast$ denotes the cyclic convolution (assuming $N$ is even due to the cryptosytem requirements).
Therefore, we can implement a generic NTT of $N$ samples with a $2N$-th root of unity $\alpha^{\frac{1}{2}}$ by simply performing the pre-processing with $\alpha^{\frac{l^2}{2}}$, convolving the pre-processed signal with $\alpha^{\frac{-l^2}{2}}$ and post-processing the convolution result with $\alpha^{\frac{l^2}{2}}$. 

\begin{figure}[ht!]
\centering
\includegraphics[width=6in]{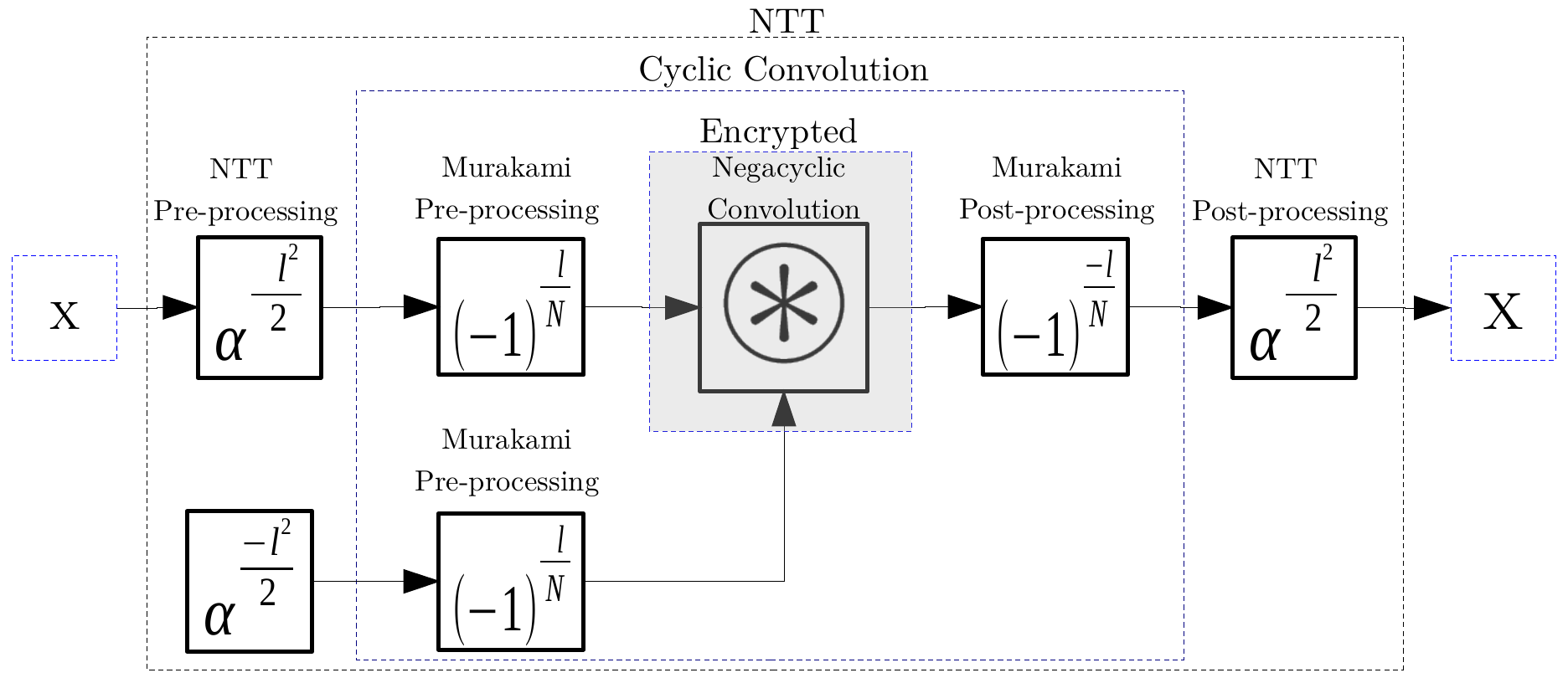}
\vspace{-0.1cm}
\caption{Block diagram for the implementation of the encrypted NTT.}
\label{fig:block_diagram}
\vspace{-0.3cm}
\end{figure}

This procedure allows to efficiently execute an encrypted NTT as shown in Figure~\ref{fig:block_diagram}. As negacyclic convolutions are the only homomorphically allowed convolutions, we resort to the pre- and post-processing shown in Section~\ref{sec:prepostproc}, which must be applied ``inside'' our convolution box (see Figure~\ref{fig:block_diagram}). Thus, $x'[l]$ is multiplied by the pre-processing vector before being encrypted. We apply the same pre-processing to $\alpha^{\frac{-l^2}{2}}$.

Finally, once the result is decrypted, we have to apply the component-wise post-processing for the cyclic convolution and, afterwards, the NTT post-processing. 

The INTT (Inverse Number Theoretic Transform) implementation is analogous to the NTT, simply swapping the used signals and including a $N^{-1}$ factor: 
\begin{align*}
  x[k] =& N^{-1}\alpha^{\frac{-k^2}{2}}\sum_{l = 0}^{N-1} \alpha^{\frac{-l^2}{2}} X[l] \alpha^{\frac{{(k - l)}^2}{2}}\\
  =& N^{-1} \alpha^{\frac{-k^2}{2}} ((\alpha^{\frac{-k^2}{2}} X[k]) \circledast \alpha^{\frac{k^2}{2}}).
\end{align*} 

A new application enabled by encrypted NTT calculations is the element-wise signal multiplication. This is accomplished by simply leveraging the cyclic convolution property of the NTT to implement point-wise products as homomorphically allowed convolutions. Consequently, we obtain the desired product with an INTT of the decrypted result. While this could also be achieved with the DFT \cite{BPB09a}, the use of NTT avoids rounding and blow-up problems under encryption.

\paragraph{Performance evaluation of the encrypted NTT}
\label{eval2}
Prior works have proposed the use of the Paillier cryptosystem for performing the DFT \cite{BPB09a}. Our method would require a multiplication step of the encrypted signal samples with powers of the corresponding $N$-th root of unity (see Section~\ref{sec:NTTs}), which cannot be encrypted due to the limited homomorphism of Paillier. However, the security of Paillier relies on the hardness of computing $\phi(N)$ (Euler's totient function) without knowing the factorization of $N$. Of course, when the different powers of the $N$-th root of unity are known, $\phi(N)$ is disclosed.

As a consequence, Paillier cannot be used for calculating the NTT without resorting to a two-party protocol for secure multiplication~\cite{CDN01} along with the corresponding increase of the execution overhead. Instead of Paillier, other schemes for which knowing the different powers of the $N$-th root of unity is not a security problem could be used (e.g., exponential El Gamal \cite{CGS97}), but they present additional drawbacks.

For this reason, we compare the efficiency of our proposed encrypted NTT with a straightforward encrypted realization of Eq.~\eqref{eq:ntt}, in which one ciphertext multiplication is used for each output NTT coefficient.
We use our aforementioned implementation of Lauter \cite{gmplib,ABGGKL16} for comparing the runtimes of the different schemes. Figures \ref{fig:exp12}c and \ref{fig:exp12}d compare the encrypted NTT performance with the straightforward application of Eq.~\eqref{eq:ntt} and our proposed method, both with Lauter ($t = F_4 = 65537$ \footnote{$F_4$ is the fifth Fermat number, where Fermat numbers are defined as the set of numbers satisfying $F_l = 2^{2^l} + 1$ with $l \in \mathbb{N}$ including zero.}). Our method enables the computation of the NTT with only one ciphertext multiplication instead of $N$, so the computational complexity is reduced in a factor of $N$.

Regarding the security, as we fix $n = N$, when we increase the length of the signals involved in the computation we also increase the achieved security (see Section~\ref{sec:security}).

Thanks to one of the anonymous reviewers, we were made aware of a work by Dor\"{o}z~\emph{et al.}~\cite{DSC15} developed in parallel and independently of our work; the authors of~\cite{DSC15} also homomorphically perform an NTT under encryption, exemplified under the LTV cryptosystem~\cite{LTV13}.
The main solution proposed in \cite{DSC15} takes advantage of a clever packing of the signals to encode each element of the original signal in different ciphertexts, and compute the corresponding fast algorithm for the NTT, enventually having the output of the NTT in $N$ different ciphertexts (one per coefficient). In order to improve the throughput they resort to batching, hence performing $N$ parallel NTTs. The computational cost of their algorithm is equivalent to $\mathcal{O}(N^2 \log_2{N})$ elemental integer multiplications between ciphertext coefficients. Compared to our scheme,\footnote{Our algorithms and those in \cite{DSC15} are exemplified in different cryptosystems, but can be independently applied to LTV or Lauter, so we find it fairer to compare their theoretical computational costs in terms of elemental operations between ciphertext coefficients instead of implementation-dependent runtimes.} it can be seen that we achieve the same computational cost; i.e., for one NTT we use one ciphertext product or $\mathcal{O}(N \log_2{N})$ elemental multiplications of coefficients. However, our solution presents two main advantages: a) it is more flexible, as we do not have to pack several messages into one ciphertext and do not require packing/unpacking operations, and b) our scheme only requires one homomorphic multiplication, while their solution requires $\log_2 N$ chained products over the same ciphertext, with the corresponding increase in ciphertext noise, in the required coefficient bitsize, and in complexity of the elemental operations, which our scheme does not incur.

If we compare the coefficient bitsize of our scheme $\log_2{q_{1}}$ with respect to Dor\"{o}z's $\log_2{q_{2}}$, it can be shown that $\left( \log_2{q_{2}} - \log_2{q_{1}} \right) \approx B\log_2{N} - B$ bits, with $B > 0$ being a constant that depends on the cryptosystem parameters. Hence, our scheme is also more efficient in terms of coefficient size (cipher expansion), by a factor of $\log_2 N$. It must be noted that this also holds for a leveled cryptosystem, where their solution would require more levels and a deeper key chain.

The encrypted fast transform is always less space-efficient (due to the need of bigger $q$) than the direct implementation, but depending on the cost of the homomorphic products, the fast algorithm can be also less time-efficient than the na\"{i}ve direct implementation due to the growth in plaintext size (accumulated quantization factors) that the former imposes, produced by its subsequent multiplications on the same ciphers ($\log_2 N$ levels), whereas the direct implementation only multiplies each cipher once. While this does not happen to Paillier~\cite{BPB09a}, it is true for lattice-based cryptosystems. Therefore, in order to mitigate this effect in their work, Dor\"oz~\emph{et al.} propose to implement the matrix multiplication associated to the NTT transform (direct transform) instead of the fast algorithm; hence, as their procedure only supports one multiplication with a cleartext constant, their cost to perform $N$ parallel NTTs is $\mathcal{O}(N^3)$ elemental multiplications between coefficients. This is considerably outperformed by our solution.

\subsection{Generalized Convolutions and Transforms} 
We can generalize the two prior primitives by adding the new pre- and post-processing and fixing one of the convolved signals, in such a way that we can achieve an INTT or NTT with any convolution type considered by Eq.~\eqref{eq:generconv}. We can formulate this as a generalization of the Murakami scheme. The general matrix scheme is as follows: 

\begin{equation*}
\bm{X} = \bm{P}_{out}\bm{G}_{\beta} \left( \bm{P}_{in,y} \bm{y} \right) \bm{P}_{in,x} \bm{x},
\end{equation*}
with $\bm{P}_{out} = \bm{P}_2(\gamma^{\frac{-1}{2}}) \bm{P}_1(\beta^{\frac{-1}{N}})$, $\bm{P}_{in,x} = \bm{P}_1(\beta^{\frac{1}{N}}) \bm{P}_2(\gamma^{\frac{-1}{2}})$ and $\bm{P}_{in,y} = \bm{P}_1(\beta^{\frac{1}{N}}) \bm{P}_2(\gamma^{\frac{1}{2}})$; and where $\bm{x}$ and $\bm{y}$ are the two input vectors, and $\bm{X}$ represents the result vector. 
$\bm{P}_i(x)$ is the following matrix

\begin{equation*}
\bm{P}_i(x) = 
\left(
\begin{array}{cccc} 
1 & 0 & \cdots & 0 \\ 
0 & x & \cdots & 0 \\ 
\vdots & \vdots & \ddots & \vdots \\ 
0 & 0 & \cdots & x^{{(N-1)}^i} 
\end{array} 
\right),
\end{equation*}
and $\bm{G}_{\beta}(\bm{v})$ is the $\beta$-generalized cyclic matrix of the vector $\bm{v}$

\begin{equation*}
\bm{G}_{\beta}(\bm{v}) = 
\left(
\begin{array}{cccc} 
v_0 & \beta v_{N-1} & \cdots & \beta v_1 \\ 
v_1 & v_0 & \cdots & \beta v_2 \\ 
\vdots & \vdots & \ddots & \vdots \\ 
v_{N-1} & v_{N-2} & \cdots & v_0 
\end{array} 
\right).
\end{equation*}

The values for the different parameters depend on the choice of convolution or transform. In the case of cyclic convolutions with our cryptosystem (only negacyclic convolutions can be homomorphically performed), we consider $\gamma^{\frac{1}{2}} = 1$ and $\beta = -1$ (only Murakami pre-/post-processing is applied). The elements $x_i$ and $y_i$, for an integer $i$ such that $0 \leq i < N$, correspond to the samples of the signals we want to convolve.

Our NTT implementation uses $\gamma = \alpha^{-1}$, $\beta = -1$ and the elements $y_i$ are equal to $1$ for all $i$. On the other hand, an INTT would use $\gamma = \alpha$, $\beta = -1$, $y_i = 1$ for all $i$, and add a multiplication by $N^{-1}$ as a post-processing step.

It must be noted that due to the requirements and structure of our cryptosystem, we implement the cyclic convolution with underlying negacyclic convolutions (i.e., we use $\beta = -1$). Conversely, it would be also possible to obtain a cyclic convolution, NTT or INTT by any other convolution type covered by Eq.~\eqref{eq:generconv} by simply using a different value of $\beta$.

\section{Optimizations}
\label{sec:optimizations}
This section presents a series of optimizations to the contributions of Section~\ref{sec:generalizeETs}, targeted at: a) efficiently performing the encrypted NTT operation by means of a relinearization primitive, b) enabling component-wise encrypted products avoiding the pre- and post-processing needed for the encrypted NTT, therefore removing the need of interaction by the secret key owner for performing an encrypted NTT, and c) enabling batch processing and maximizing the batched homomorphic capacity. For these purposes, we rely on a relinearization step and the CRT (Chinese Remainder Theorem), and we exploit the periodic structure of the input signals whenever they present it. We first revise the formulation of the relinearization primitive, and then explain the proposed optimizations.

\subsection{Relinearization primitive}
\label{sec:relinearization}
For the purpose of avoiding pre- and post-processing in the cleartexts, we can employ a \emph{relinearization} primitive \cite{LNV11,BV11aJ,BV11relinearization}, commonly used in key switching algorithms to reduce the size of the encryptions after a multiplication: when multiplying two ciphertexts $\bm{c} = (c_0, c_1)$ and $\bm{c}'= (c'_0, c'_1)$ from the cryptosystems \cite{LNV11}, \cite{BV11b} and \cite{PTP15}, the number of elements of the resulting ciphertext is increased $\bm{c}'' = (c''_0, c''_1, c''_2)$. Hence, $\bm{c}''$ could be decrypted as $c''_0 + c''_1s + c''_2s^2$, which can be seen as a quadratic function of $s$.

This increase is undesired due to the induced overheads. Hence, a relinearization reduces $\bm{c}''$ to a new ciphertext formed by only two elements $\bm{c}''' = (c'''_0,c'''_1)$, satisfying $D(\bm{c}''',s) = D(\bm{c}'',s)$, where $D(\bm{c},s)$ represents the decryption of $\bm{c}$ with key $s$ (the decryption circuits for both cases are $c'''_0 + c'''_1s$ and $c''_0 + c''_1s + c''_2s^2$, respectively).
In order to perform this relinearization, the public key must comprise certain additional information about the successive powers of $s$, and circular security must hold for the cryptosystem to securely encrypt functions of the secret key. In case of applying the relinearization after each product, only information of $s^2$ is needed. As a drawback, the relinearization increases the ciphertext noise.

This is the conventional use of relinearization, but we use it additionally for performing other types of operations as upsampling, downsampling and reflections (see Section~\ref{sec:applications}). 
We present now the formulation of a relinearization step; in our work we do not restrict the relinearization to only powers of the secret key $s$, and we consider a generic decryption for a ciphertext $(c_0, c_1, c_2)$ as $c_0 + c_1s_1 + c_2s_2$, with $s_1,s_2 \leftarrow \chi$, where $s_2$ is not necessarily equal to $s_1^2$. It can be shown that the needed additional information would be $(h_1, \dots, h_{\lceil \log_t q \rceil -1})$, where the $h_i$ are \emph{key homomorphisms} defined as
\begin{equation*} 
h_i = (a_i, b_i = -(a_is_1 + te_i) + t^is_2),\; i = 0, \dots, \lceil \log_t q \rceil -1,
\end{equation*} 
where $t$ is the module used for encoding the messages in $R_t$, $a_i \leftarrow R_q$ and $e_i \leftarrow \chi$. Expressing $c_2$ in base-$t$ representation, we have $c_2 = \sum_{i=0}^{\lceil \log_t q \rceil - 1} c_{2,i} t^i$, and finally we obtain the ciphertext $(c_0^{\textit{relin}},c_1^{\textit{relin}})$ under the key $s_1$ as 
\begin{equation*}  
c_0^{\textit{relin}} = c_0 + \sum_{i=0}^{\lceil \log_t q \rceil - 1} c_{2,i} b_i, \quad
c_1^{\textit{relin}} = c_1 + \sum_{i=0}^{\lceil \log_t q \rceil - 1} c_{2,i} a_i.
\end{equation*} 

This step can be typically used either after each encrypted multiplication, in order to bring back the expanded ciphertext to a pair of polynomials, or after several consecutive multiplications, by using relinearizations $h_i$ for each different key power. In general, given a ciphertext $(c_0, c_1, \ldots, c_{m-1})$, whose decryption function has the form $c_0 + \sum_{i=1}^{m-1} c_is_i$, we can implement $m-2$ relinearizations to convert the ciphertext into a linear equation in terms of a unique key; e.g., if we want to express all polynomials as a function of key $s_1$, we use $m-2$ key homomorphisms $h^{(j)}_i$, where each homomorphism would have the key $s_j$ ``encrypted'' in terms of $s_1$ for $j = 2, 3, \ldots, m-1$.
By recursively applying these relinearizations, i.e. $h^{(2)}_i$ to the ciphertext composed by $(c_0, c_1, c_2)$, then $h^{(3)}_i$ to the concatenation of the previous result and $c_3$, and so on, we arrive at the expression that encompasses all concatenated relinearizations in two equations:
\begin{align*} 
c_0^{\textit{relin}} =& c_0 + \sum_{i=0}^{\lceil \log_t q \rceil - 1} c_{2,i} b_i^{(2)} + \ldots + \sum_{i=0}^{\lceil \log_t q \rceil - 1} c_{m,i} b_i^{(m)}, \\
c_1^{\textit{relin}} =& c_1 + \sum_{i=0}^{\lceil \log_t q \rceil - 1} c_{2,i} a_i^{(2)} + \ldots + \sum_{i=0}^{\lceil \log_t q \rceil - 1} c_{m,i} a_i^{(m)}.
\end{align*}

Now, considering the vectors
\begin{align*}
  \bm{c}_{\textit{base}-t} =& {(\bm{c}_{2,\textit{base}-t}^T, \bm{c}_{3,\textit{base}-t}^T, \ldots, \bm{c}_{m,\textit{base}-t}^T)}^T, \\ 
  \bm{a} =& {\left( {\left(\bm{a}^{(2)} \right)}^T, {\left(\bm{a}^{(3)} \right)}^T, \ldots, {\left(\bm{a}^{(m)} \right)}^T \right)}^T, \\
  \bm{b} =& {\left( {\left(\bm{b}^{(2)} \right)}^T, {\left(\bm{b}^{(3)} \right)}^T, \ldots, {\left(\bm{b}^{(m)} \right)}^T \right)}^T,
\end{align*}
where
\begin{equation*}
\begin{split}
 & \bm{c}_{i,\textit{base}-t} = {(c_{i,0}, c_{i,1}, \ldots, c_{i,\lceil \log_t q \rceil - 1})}^T, \\ 
 & \bm{a}^{(i)} = {(a_{0}^{(i)} , a_{1}^{(i)}, \ldots, a_{\lceil \log_t q \rceil - 1}^{(i)})}^T, \\
 & \bm{b}^{(i)} = {(b_{0}^{(i)}, b_{1}^{(i)}, \ldots, b_{\lceil \log_t q \rceil - 1}^{(i)})}^T,
\end{split}
\end{equation*}
we get the simplified vector expression of the relinearization
\begin{equation*} 
c_0^{\textit{relin}} = c_0 + \bm{c}_{\textit{base}-t} \cdot \bm{b},\quad
c_1^{\textit{relin}} = c_1 + \bm{c}_{\textit{base}-t} \cdot \bm{a},
\end{equation*}
where $\bm{a} \cdot \bm{b}$ is the scalar product between the vectors of polynomials $\bm{a}$ and $\bm{b}$. With this expression, if the key owner generates the appropriate relinearization matrices, we can flexibly convert encryptions between different keys (key switching) and even extract or linearly combine different individual components of an encrypted polynomial signal.

\subsubsection{Increase of error after relinearization}

The noise added to the ciphertext after the execution of one relinearization step is approximately equivalent to the noise added over the same ciphertext by as many homomomorphic additions of fresh ciphertexts as polynomials compose the vectors $\bm{a}$ and $\bm{b}$. Therefore, if both vectors have $m \lceil \log_t{q} \rceil$ polynomial elements, it is equivalent to $m \lceil \log_t{q} \rceil$ homomorphic additions. Hence, even when some of the proposed methods in this work resort to a relinearization step, they still allow for $\mathcal{O}(D)$ homomorphic products between ciphertexts, being $D$ the number of products allowed by the choice of $q$ (see Eq.~\eqref{eq:condq}).
 
\subsection{Proposed Optimizations}
\label{optimizations}
This section introduces several strategies based on relinearization, aimed at optimizing the realization of the encrypted NTT proposed in Section~\ref{sec:encryptedNTTs}, avoiding the interaction with the secret-key owner for the pre- and post-processing steps; for this purpose, we take advantage of the specific structure of the transform matrices. We first present a polyphase-decomposition-based approach which reduces the key size, and then enhance it by preserving the key security. The polyphase decomposition is a common tool used in signal processing \cite{Harris04}, which has also been applied in a cryptographic setting as a means to achieve different tradeoffs in the General Learning with Errors (GLWE) problem \cite{BV11aJ}.

Our target is to calculate the NTT of an already encrypted version of $x[l]$, which has not been pre-processed (Section~\ref{sec:prepostNTTs}). We first note that our encrypted NTT algorithm allows to perform one of the processings under encryption, by expressing it as a convolution. For the NTT we have
\begin{equation*} 
\mathrm{NTT}(x[l]) = N^{-1}\left( \left( x[k] \circledast \alpha^{\frac{-k^2}{2}} \right) \alpha^{\frac{k^2}{2}} \right) \circledast \mathrm{NTT} \left( \alpha^{\frac{l^2}{2}} \right).
\end{equation*} 

Analogously, for the INTT we have
\begin{equation*} 
\mathrm{INTT}(x[l]) = N^{-1}\left( \left( x[k] \circledast \alpha^{\frac{k^2}{2}} \right) \alpha^{\frac{-k^2}{2}} \right) \circledast \mathrm{INTT} \left( \alpha^{\frac{-l^2}{2}} \right).
\end{equation*} 

With this structure, we only have to implement \emph{one of the component-wise products} with the (known) pre- or post-processing vector under encryption to get an unattended implementation of the encrypted NTT; we use the polyphase decomposition of the inputs and exploit the use of the relinearization to homomorphically calculate the pre- or post-processing.

\subsubsection{Encrypted NTT with polyphase decompositions}
\label{polyphase-decomposition}
In order to calculate a component-wise product of the signals $x[l]$ and $h[l]$, we can decompose them in as many polyphase components as their length. A relinearization can be used to extract each of these components into separate encryptions, and subsequently, element-wise multiplication can be straightforwardly performed. Then, an inverse relinearization step would enable regrouping the signals in a sole encryption which can be decrypted using the initial secret key.
This approach suffers from an excessive computational cost to perform $N$ relinearizations; moreover, the corresponding relinearizations to one polyphase component reduce the problem to a lattice with $n = 1$, which would imply no security. 

We can exploit the use of the polyphase decomposition in a smarter way, balancing the size of the used relinearization matrices and the reduction in the security: We divide the signal in a number of components equal to a constant $M$ (the previous solution corresponds to $M = N$). For our choice of cryptosystem parameters (see Section \ref{sec:prelims}), we need $M$ be a power of two dividing $N$.
Hence, we can express the element-wise multiplication in terms of $M$ smaller and independent homomorphic element-wise multiplications, where each signal has a size $N/M$. Therefore, we are able to divide the sought encrypted operations as a set of simpler and easier element-wise operations (with the corresponding reduction in the considered lattice). This process could be recursively performed, at the cost of eventually reducing again the ciphertexts to a lattice with $n = 1$, which is unacceptable in terms of security.

We still need a method to homomorphically perform the element-wise operations without resorting to a reduction in the dimension of the lattice, which is presented in detail in Section~\ref{element-wise-processing}. By combining this method with the partial polyphase decomposition we can produce several possible solutions for an encrypted element-wise multiplication. Depending on the chosen $M$, we can trade-off efficiency (lower size for the relinearization matrices) for security (lower $n$).

Our proposed process for an encrypted component-wise product $x[l] \alpha^{\frac{l^2}{2}}$ is the following:
\begin{itemize}
\item Decimate $x[l + m]$ with $m \in [0, M)$ by a factor $M$.
\item For each polyphase component, apply a relinearization encrypting the corresponding component in a polynomial ring isomorphic to a lattice of dimension $\frac{N}{M}$ (if $M > 1$).
\item Perform the element-wise multiplication of each component by the corresponding component of the signal $\alpha^{\frac{l^2}{2}}$ by resorting to the method proposed in Section~\ref{element-wise-processing} (multiplication between a ciphertext and a cleartext). If $M = N$, the multiplication can be directly performed.
\item Finally, a reverse relinearization process is applied to each component so that they are regrouped into a new ciphertext under the same key (if $M > 1$).
\end{itemize}

This method produces an element-wise product by the pre- or post-processing vector without the intervention of the secret key owner, enabling a fully non-interactive computation of the encrypted NTT. Moreover, the used relinearization decreases the dimension by a factor $M$ (each polyphase component has a length of $\frac{N}{M}$ samples), thereby achieving a net improvement in both computational cost and security with respect to a direct application of the polyphase decomposition ($M = N$).

\subsubsection{Encrypted NTT without lattice dimension reduction}
Decreasing the size of the used key as done by the previous method implies a reduction in security.
However, it is possible to perform the sought element-wise multiplication between a ciphertext and a known cleartext vector with no such reduction. Hence, we enable additional secure and efficient operations like modulation or demodulation with an unencrypted carrier, or the implementation of the encrypted NTT without the intervention of the key owner, which is our purpose.

First, we show how to perform the element-wise multiplication of a ciphertext and cleartext without a reduction in the lattice dimension. Finally, we explain how to use smaller relinearizations and achieve a net improvement in the efficiency of the operations when the cleartext is periodic.

\paragraph{Element-wise product between ciphertext and cleartext}
\label{element-wise-processing}
We consider the ciphertext $\bm{c} = (c_0, c_1)$, whose decryption circuit would be $c_0 + c_1s$, and the polynomial represented as a column vector $\bm{g} = (g_0, g_1, \ldots , g_{n-1})^T$.
If we denote by the polynomial $c'_0$ the result of the element-wise multiplication between $c_0$ and $g$, the decryption circuit in matrix form will be $\bm{c}'_0 + \mbox{diag$(\bm{g})$} \bm{C_1} \bm{s}$, where $\mbox{diag$(\bm{g})$}$ is a diagonal matrix composed of the elements of the vector $\bm{g}$, and $\bm{C_1}$ is the skew circulant matrix \cite{Dav12} of the polynomial $c_1$.

Now, we apply the relinearization algorithm and express the decryption circuit in terms of polynomial products or, in matrix form, products between vectors and skew circulant matrices. Considering the key homomorphism $h_i = (a_i, b_i = -(sa_i + te_i) + t^is)$, with $i = 0, \ldots, \lceil \log_t q \rceil - 1$, we have
\begin{equation*}
\bm{c}_0^{\textit{relin}} = \bm{c}'_0 + \sum_{i = 0}^{\lceil \log_t q \rceil - 1} \bm{\tilde{C}}_{i} \bm{b}_i,\quad
\bm{c}_1^{\textit{relin}} = \sum_{i = 0}^{\lceil \log_t q \rceil - 1} \bm{\tilde{C}}_{i} \bm{a}_i,
\end{equation*}
where $\bm{\tilde{C}}_{i}$, $i = 0, \ldots, \lceil \log_t q \rceil - 1$, is the base-$t$ decomposition of the matrix product $\mbox{diag$(\bm{g})$}\bm{C}_1$. For the decryption circuit $c_0^{\textit{relin}} + c_1^{\textit{relin}}s$ to be correct, $\bm{S\tilde{C}}_{i}$ has to be equal to $\bm{\tilde{C}}_{i}\bm{S}$ for all $i$, with $\bm{S}$ being the negacyclic or skew circulant matrix corresponding to the polynomial $s$. The previous equality is true when all the $g_i$ are equal (multiplication by a polynomial of maximum degree $0$), but in our general case all the $g_i$ are different, and equality is not achieved. Therefore, the ciphertext must be modified to perform the sought relinearization. $\mbox{diag$(\bm{g})$} \bm{C}_1 \bm{s}$ can be expressed equivalently in polynomial form as $\sum_{j = 0}^{n-1} c^{(j)}(z) s_j$, being $c^{(j)}(z)$ the polynomial whose coefficients are the $j$-th column of the matrix product $\mbox{diag$(\bm{g})$} \bm{C}_1$. Finally, with these requirements the new decryption circuit has the form $c'_0(z) + \sum_{j = 0}^{n-1} c^{(j)}(z)s_j$.

Now, considering $n$ key homomorphisms $h_i^{(j)}$ with $i = 0, \ldots, \lceil \log_t q \rceil - 1$ and $j = 0, \ldots, n-1$, in which $h_i^{(j)}$ has the coefficient $s_j$ ``encrypted'' under the key $s$, we can perform a unique relinearization by concatenating all the $h_i^{(j)}$ and the corresponding polynomials $c^{(j)}(z)$ as discussed in Section~\ref{sec:relinearization}. 
As we targeted, the proposed relinearization does not convey a reduction in the size $n$ of the lattice. Regarding the computational cost of the approach, the key owner needs to generate the vectors $\bm{a}$, $\bm{b}$ and $\bm{c}_{\textit{base}-t}$ of size $n \lceil \log_t q \rceil$, which are composed of polynomials of $n$ coefficients. The relinearization comprises one polynomial addition and two scalar products $\bm{c}_{\textit{base}-t} \cdot \bm{b}$ and $\bm{c}_{\textit{base}-t} \cdot \bm{a}$, i.e, $2n \lceil \log_t q \rceil$ polynomial products and $2n \lceil \log_t q \rceil - 1$ polynomial additions.

\paragraph{Element-wise multiplication between ciphertext and periodic cleartext}
\label{periodic-element-wise-processing}
When the cleartext used in the element-wise multiplication is periodic, the length of the vectors $\bm{a}$ and $\bm{b}$ required for the relinearization process can be reduced, therefore decreasing the number of addends in the decryption circuit. If $g$ is a periodic signal with $m$ samples per period, and $m$ divides $n$, we can use the following decryption circuit:
\begin{equation*}
c'_0 + \sum_{i = 0}^{m - 1} \underbrace{z^i \left( \sum_{j = 0}^{m - 1} g_{i + j \bmod m} z^j c^{(j)}\left(z^m \right)\right)}_{c'_{i+1}}s^{(i)}\left(z^m\right),
\end{equation*}
where $s^{(i)}\left(z\right) = \sum_{d = 0}^{\frac{n}{m} - 1} s[md - i] z^d$ and $c^{(j)}\left(z\right) = \sum_{d = 0}^{\frac{n}{m} - 1} c_1[md - j] z^d$ are the $i$-th and $j$-th components of the polyphase decomposition in $m$ components of $s$ and $c_1$, respectively, and $s^{(i)}\left(z^m\right)$, $i = 0, \ldots, m-1$, are the corresponding secret keys.

The obtained ciphertext consists of $m + 1$ polynomials; applying the relinearization as in the previous section, we reduce it to only two components. Hence, when $g$ is periodic with a period of $m$ samples, it is possible to reduce the size of the vectors $\bm{a}$, $\bm{b}$ and $\bm{c}_{\textit{base}-t}$ to $m \lceil \log_t q \rceil$ components each. Regarding the complexity, the proposed relinearization requires $2m \lceil \log_t q \rceil$ polynomial products and $2m \lceil \log_t q \rceil - 1$ polynomial additions.

It is worth noting that the pre and post-processing vectors needed to implement the encrypted NTT and INTT present some additional structure and periodicities which could be exploited in order to increase the efficiency of the computation of the encrypted NTT. Additionally, the solutions presented in this section could also be useful for other typical signal processing applications involving periodic signals, like modulations and demodulations (see \ref{modulation}).

\subsection{Element-wise multiplication of two encrypted messages}
\label{element-wise-final-method}
The previous sections describe a fully non-interactive encrypted NTT with efficient relinearization operations which enable component-wise products between an encrypted vector and a known clear-text vector. We can now leverage the encrypted NTT to perform component-wise products between two fully encrypted vectors of length $N$ without reducing the security of the scheme. Fixing the parameter $n$ and working with $\lceil \frac{N}{n} \rceil$ pairs of ciphertexts, the computational cost in terms of elemental products$\mbox{}\bmod{q}$ would be $\mathcal{O}(2^{\lceil \log_2 N \rceil } n \log_2 n) \approx \mathcal{O}(N n \log_2 n)$, step-wise linear in terms of $N$.

For the sake of comparison, we define $\mbox{cost}_1$ as the computational cost of the techniques presented in Section~\ref{element-wise-processing} for computing of the encrypted NTT, $\mbox{cost}_2$ for the polyphase-decomposition-based method which with $M$ components, and $\mbox{cost}_3$ for the straightforward decomposition in $N$ components (see Section \ref{polyphase-decomposition}). Hence, we obtain the ratios $\frac{\mbox{cost}_2}{\mbox{cost}_1} \approx \left( \frac{M}{n} + \frac{1}{M} \right)\left( 1 - \frac{\log_2{M}}{\log_2{n}} \right) + \frac{1}{n}$ and $\frac{\mbox{cost}_3}{\mbox{cost}_1} \approx \frac{1}{\log_2{n}} + \frac{1}{n}$. The computational cost for the element-wise multiplication between two encrypted messages is approximately bounded by three times the computational cost for an encrypted and a clear-text message. This is due to the need of homomorphically computing two NTT and one INTT, which amounts to three executions of the algorithms (or only some parts of the algorithms) previously presented in~\ref{optimizations}. Table~\ref{tab:element-wise} summarizes the computational cost in terms of elemental products modulo $q$, the total size of the relinearization matrices in terms of coefficients modulo $q$ and the minimum required dimension for the lattices in the three methods.

\begin{table}[!t]
\renewcommand{\arraystretch}{1.3}
\caption{Comparison of the proposed encrypted element-wise multiplication methods}
\label{tab:element-wise}
\centering \scriptsize
\begin{tabular}{|c|c|c|}
\hline
\multicolumn{3}{|c|}{Computational Cost}\\
\hline 
\multicolumn{3}{|c|}{$\mbox{cost}_1 \approx \mathcal{O} \left( \lceil \frac{N}{n} \rceil n^2 \log_2{n} \right)$} \\
\multicolumn{3}{|c|}{$\mbox{cost}_2 \approx \mathcal{O} \left( \lceil \frac{N}{n} \rceil \left( \left( Mn + \frac{n^2}{M} \right)\log_2{ \left( \frac{n}{M} \right) } + n \log_2{n} \right) \right)$} \\
\multicolumn{3}{|c|}{$\mbox{cost}_3 \approx \mathcal{O} \left( \lceil \frac{N}{n} \rceil \left( n^2 + n \log_2{n}\right) \right)$} \\
\hline
\multicolumn{3}{c}{} \\
\cline{1-1} \cline{3-3}
Total size of the relin. matrices & & Minimum lattice dimension \\ \cline{1-1} \cline{3-3}
 $\mbox{bits}_{1} = 2 (n^2 + n) \lceil \log_{t}{q} \rceil \lceil \log_{2}{q} \rceil$ & & $n_{{min}_1} = n$ \\
 $\mbox{bits}_{2} = \left( 4n + \frac{2n^2}{M^2} + \frac{2n}{M} \right)\lceil \log_{t}{q} \rceil \lceil \log_{2}{q} \rceil$ & & $n_{{min}_2} = \frac{n}{M}$ \\
 $\mbox{bits}_{3} = (4n + 2) \lceil \log_{t}{q} \rceil \lceil \log_{2}{q} \rceil$ & & $n_{{min}_3} = 1$ \\ \cline{1-1} \cline{3-3}
\end{tabular}
\vspace{-0.2cm}
\end{table}

By increasing the parameter $M$ we achieve a net improvement in both the size of the vectors $\bm{a}$ and $\bm{b}$, and the efficiency of the encrypted element-wise multiplication, at the cost of a reduction in the underlying dimension of the lattice. This trade-off between the security and the implementation runtimes is quantified in the next section.

\subsubsection{Performance evaluation of the element-wise multiplication}
\label{eval3}
The proposed constructions enable component-wise processing in RLWE cryptosystems, which are suited and very efficient for polynomial processing. Therefore, we compare our proposed methods with the use of a cryptosystem (Paillier) which is apparently more amenable to element-wise multiplication than RLWE-based cryptosystems. We use Lauter's cryptosystem to implement our proposed methods for element-wise products (Sections~\ref{polyphase-decomposition}, \ref{element-wise-processing} and \ref{element-wise-final-method}).

In general, the computational cost for performing $N$ element-wise multiplications with Paillier (with one of the messages in clear) is $N$ modular exponentiations of Paillier ciphertexts. With Lauter, we would need only $N$ element-wise multiplications (see Section~\ref{element-wise-processing}), but the computational cost for the relinearization is relatively high.
In order to have a fair comparison, we fix a value of $n$ (polynomial degree) independent of $N$ (message size) for Lauter, as a function of the needed level of security. Then, the computational cost for the element-wise multiplication of two pairs of $N$ encrypted integers would be approximately $\mathcal{O} \left(\lceil \frac{N}{n} \rceil n^2\log_2n \right) = \mathcal{O} \left(2^{\lceil \log_2 N \rceil} n\log_2n \right) \approx \mathcal{O} \left(Nn \log_2 n \right)$ for the techniques from Section~\ref{element-wise-processing} or $\mathcal{O}(2^{\lceil \log_2 N \rceil } \left( M + \frac{n}{M} \right) \log_2{\left( \frac{n}{M} \right)} + \log_2{n}) \approx \mathcal{O} \left( \left( NM + \frac{Nn}{M} \right) \log_2{\left( \frac{n}{M} \right)} + N\log_2{n} \right)$ when resorting to polyphase decompositions (Section~\ref{polyphase-decomposition}), using for both a radix-2 algorithm (as we have already described in the previous section \ref{element-wise-final-method}).

\begin{figure}[t]
\centering
\includegraphics[height=3.5in,width=6in]{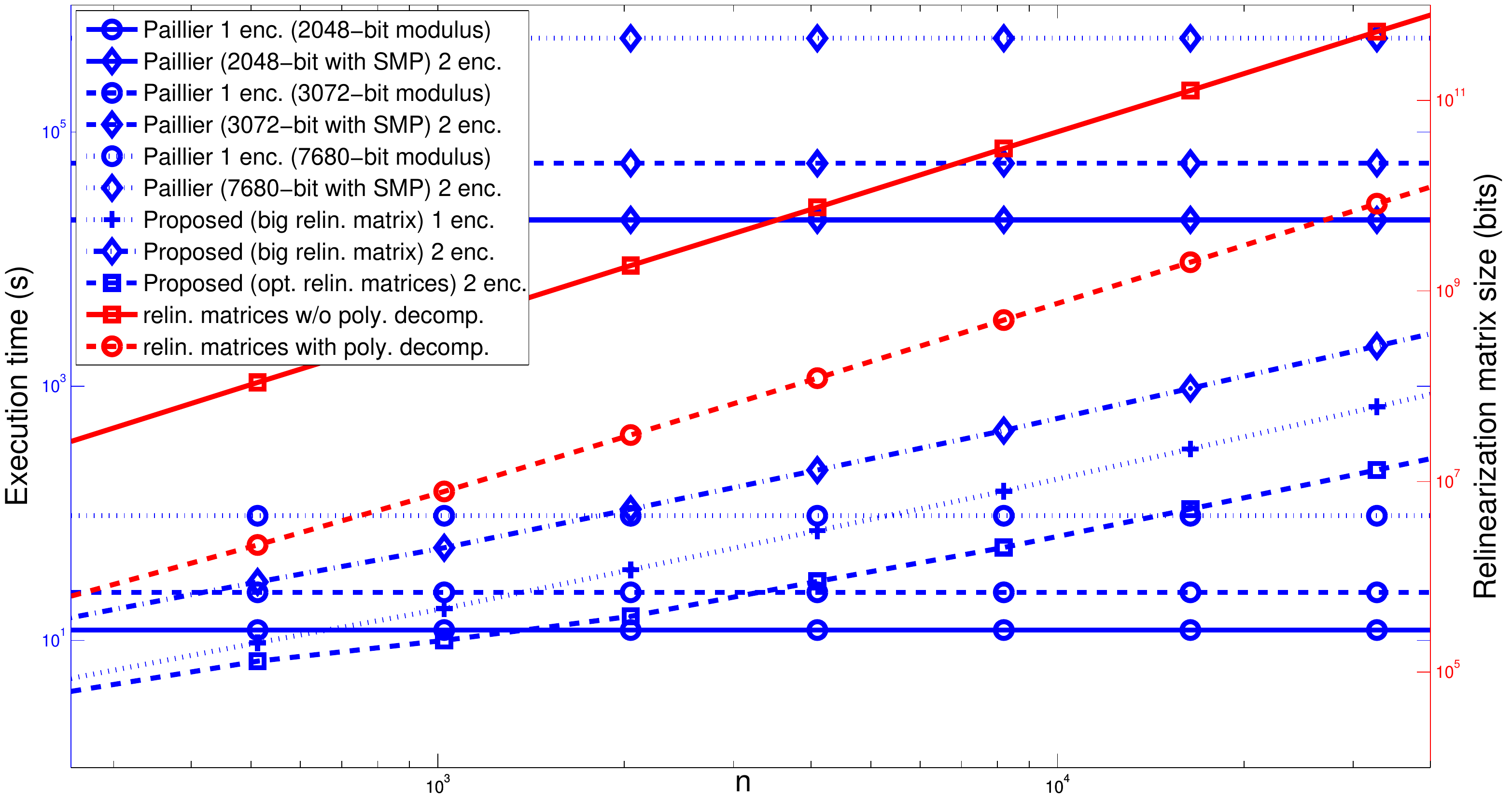}
\vspace*{-2mm}
\caption{Comparison of element-wise runtimes for different schemes.}
\label{fig:element-wise}
\vspace{-0.3cm}
\end{figure}

Figure~\ref{fig:element-wise} compares the different runtimes for a) the element-wise method in Section~\ref{element-wise-processing}, b) the polyphase-based method from Section~\ref{element-wise-final-method}, c) the partial polyphase method from Section~\ref{element-wise-final-method} with $M = 8$, d) Paillier-based element-wise multiplication between an encrypted message and a message in the clear, and e) the combination of Paillier-encrypted messages and a secure interactive multiplication protocol (SMP) \cite{CDN01} for computing the product of two encrypted messages. In all the experiments, we have chosen practical parameters for the Paillier cryptosystem ($2048$-bit, $3072$-bit and $7680$-bit moduli) and we vary the $n$ used for the lattice cryptosystems. We are considering $N = 131072$ (the runtimes increase linearly with $N$). 
Additionally, Figure~\ref{fig:element-wise} also compares the bit-size of the relinearization matrices used for the first and second methods from Section~\ref{element-wise-final-method} in terms of $n$.

As mentioned in previous sections, increasing $n$ produces a smaller $\delta$, and therefore a higher security (see Section~\ref{sec:security}). Hence, depending on the required security for the applications, we can choose an adequate value for $n$.

We can see that using practical values for both RLWE cryptosystems (e.g., $n = 2048$) and Paillier, the proposed optimized methods for element-wise multiplications of two encrypted vectors outperform the other approaches in terms of efficiency. In fact, Paillier can only achieve better performance when one of the vectors is unencrypted, but even in such case, RLWE-based cryptosystems are still much more efficient for polynomial operations and, when combined with our methods, they provide greater flexibility and a full toolset of unattended efficient encrypted operations along with these element-wise operations, which Paillier could not provide.

\subsection{CRT for cleartext batching}
The second optimization we propose deals with cleartext batching and enhancing the homomorphic capacity when SIMD (Single-Instruction-Multiple-Data) operations are implemented. For this purpose, we resort to the Chinese Remainder Theorem (CRT). The CRT has been used in numerous different applications, ranging from the conversion of a one-dimensional convolution into a convolution with smaller multidimensional signals, to the development of error correcting codes, secret-sharing and many more \cite{DPS96}. 

We first revisit the CRT with a notation slightly adapted to our particular scheme. We begin with the rings $R_{t_i^{k_i}}[z] = \mathbb{Z}_{t_i^{k_i}}[z] / f(z)$ and polynomials $a_i \leftarrow R_{t_i^{k_i}}[z]$, with $i = 1, \ldots, m$. If $a_i \equiv a_j \bmod{(\gcd(t_i^{k_i}, t_j^{k_j}))}$ holds for $1 \leq i, j \leq m$, then there exists a polynomial $a \in R_t[z]$ with $R_t[z] = \mathbb{Z}_{t}[z] / f(z)$ and $t = t_1^{k_1} t_2^{k_2} \ldots t_m^{k_m}$ which verifies: 
\begin{equation}
a \equiv a_i \bmod{({t_i}^{k_i})}, \mbox{ for $i = 1, \ldots, m.$}\label{eq:congruences}
\end{equation} 

For the existence of $a$, we can impose a less demanding requirement: it suffices that the $t_i$ be pairwise coprime, i.e., $\gcd(t_i, t_j) = 1$ with $i \neq j$. 
In order to find the polynomial $a\in R_t[z]$ that satisfies the above congruences \eqref{eq:congruences}, we write
\begin{equation*} 
a = \sum_{i = 1}^{m} T_i a_i d_i,
\end{equation*} 
where $T_i = t / t_i ^ {k_i}$ and $d_i$ fulfill $d_iT_i \equiv 1 \bmod{(t_i^{k_i})}$.

Therefore, the existing isomorphism between $R_t[z]$ and $R_{t_1^{k_1}}[z] \oplus R_{t_2^{k_2}}[z] \oplus \dots \oplus R_{t_m^{k_m}}[z]$ enables several cleartext operations through a single encrypted homomorphic operation. The possibility of exploiting this isomorphism to parallelize cleartext operations has been previously suggested by several authors \cite{BV11aJ, SV14}. Smart and Vercauteren~\cite{SV14} propose and exemplify SIMD operations using FHE (Fully Homomorphic Encryption) cryptosystems, for performing AES encryption homomorphically and for searching in an encrypted database. Brakerski \emph{et al.}~\cite{BV11aJ} propose batching the bootstrapping operation for improving the efficiency of the cryptosystem.

Our contribution comprises choosing appropriate values for $t$, such that the CRT can be applied to parallelize any of the encrypted operations introduced in the previous sections.

\subsubsection{Throughput optimizations for signal processing applications}
In general, prior works dealing with batching operations (see~\cite{LN14} for a comparison) are mainly focused in maximizing the throughput of operations, but generally overlook the type and usefulness of the parallelized encrypted operations, which might be severely affected by the decomposition of the cryptosystem ring in unequal prime ideals.
Contrarily, we want to present the use of the NTT as a tool for batching operations which can be more suitable for typical signal processing applications. Then, following the steps presented in Section~\ref{sec:generalizeETs}, we briefly explain how to achieve the maximum number of parallel operations between either integers or discrete signals, keeping the meaning and usefulness of the batched operations.

We assume that the used modular function is $f(z) = 1 + z^n$ and we also assume, without loss of generality, that all the $t_i$ from Eq.~\eqref{eq:congruences} are different prime numbers. Then, for the previously introduced ring $R_t$, an addition or multiplication between two ciphertexts actually conveys the element-wise addition or multiplication between the vectors whose $i$-th element belongs to $R_{t_i^{k_i}}[z]$. In this case, if we want to perform the maximum number of parallel multiplications between integers, we have to restrict the input signals to zero-degree polynomials, therefore wasting much of the plaintext space.

By resorting to the proposed pre-processing techniques and the use of the NTT, we can fully utilize all the available plaintext space. That is, combining the CRT and the NTT we can perform multiplications among integers belonging to the finite field $\mathbb{Z}_{t_i^{k_i}}$. In any case, we can maximize the number of encoded integers if we choose the right values for the different $t_i$ and $k_i$, and we show how in the following discussion.

There exists an isomorphism between the finite field $\mathbb{Z}_{t_i^{k_i}}$ and $\mathbb{Z}_{t_i}[x]/g(x)$ where $g(x)$ is an irreducible function over $\mathbb{Z}_{t_i}$. Then, in order to have $g(x) = 1 + x^{k_i}$ ($1 - x^{k_i}$ is not irreducible), we must use a cyclotomic polynomial $\Phi_{2k_i}(x) = 1 + x^{k_i}$ (with $k_i$ a power of two) and, finally, we can assert that $\Phi_{2k_i}(x)$ is irreducible over $\mathbb{Z}_{t_i}$ when it satisfies
\begin{equation}
t_i^{\phi(2k_i)} = t_i^{k_i} \equiv 1 \bmod{ 2k_i},\label{eq:crt}
\end{equation}
where $\phi(\cdot)$ is Euler's totient function and $k_i$ is the smallest integer satisfying the above condition, that is, $t_i$ is a generator of the multiplicative group $\mathbb{Z}^{*}_{2k_i}$. Further details about working with finite fields can be found in~\cite{LiNi94}.

As an example, if we consider $k_i = 2$ for all $i$, we know that if $t_i = 3 \bmod 4$, then $\mathbb{Z}_{t_i^2}$ is equivalent to $\mathbb{Z}_{t_i}[x]/(1 + x^2)$. Therefore, reusing again the proposed pre-processing and the NTT over each $\mathbb{Z}_{t_i}[x]/(1 + x^{k_i})$ we can implement the batched multiplication of more integers. 
Unfortunately~\cite{Brett07}, the only cyclotomic polynomial that allows to encode more integers while satisfying our requirements is $\phi_{4}(x)$; e.g., if we have two messages $x_1,x_2 \in R_t[z]$ and each one encodes $2mn$ integers ($2n$ integers over $\mathbb{Z}_{t_i}$ for $i = 1,\ldots,m$) thanks to the CRT, we can apply the proposed pre-processing for performing a cyclic convolution and, afterwards, use the NTT for performing the element-wise product as a cyclic convolution (first, for distinguishing the different rings $R_{t_i^{2}}$, and afterwards, for distinguishing the two different $\mathbb{Z}_{t_i}$ which we can find on the finite field $\mathbb{Z}_{t_i^{2}}$). In this way, with only one multiplication between two ciphertexts we can perfom the element-wise multiplication between $2mn$ pairs of integers. Conversely, the element-wise addition is easily obtained without the need of the NTT or pre-processing. 
In case we want to perform linear filterings instead of multiplications of integers, we can use the techniques presented in~\cite{TKCL07, BPB10} for packing a different signal in each of the involved integers.

\section{Applications: Encrypted Signal Processing Toolset}
\label{sec:applications}
This section exemplifies the use of the primitives and algorithms presented in previous sections by proposing a set of practical tools and applications, which comprise matrix operations, Cyclic Redundancy Check (CRC) codes, changes in sampling rate and linear transforms; we also show how they can be seamlessly adopted within any RLWE-based cryptosystem \cite{LNV11,BV11b}, by taking advantage of its polynomial structure. For simplicity, we assume that all signals have only one independent variable (univariate) but the methods can be easily extended to the multivariate case \cite{PTP15}.

\subsection{Typical Operations in Signal Processing} 
We present efficient methods to implement elementary signal processing operations in the encrypted domain when using lattice-based cryptosystems. Besides the different types of convolutions tackled above, shifts and scaling of the independent variable of the signals are also very common operations in signal processing. In general, shift operations do not involve any change in the cryptosystem parameters, but this is not true for operations that cause a change in the sampling rate of the encrypted signal. In that case, it is necessary to ``reset'' the secret key to the new sampling rate of the signal. Below, we address shift operations and changes in sampling rate together with modulation and demodulation operations which are enabled by using different types of relinearizations. 

\subsubsection{Shift} 
A shift $x[l-l_0]$ represents the signal $x[l]$ delayed by $l_0$ samples. This operation can be implemented on encrypted signals, represented as $z$-transform polynomials, by simply multiplying them by the monomial $z^{l_0}$. Therefore, the cost of the operation is the product of a single polynomial. Also, if the polynomial $z^{l_0}$ is available in the clear, the cost is much lower, since it only involves a product by $z^{l_0}$ in the clear with modular function $f(z)$. In case the shift makes the signal wrap-around, the same effects explained for the $\alpha$-generalized convolution would apply, and pre- and post-procesing can be used to preserve the desired sign for the wrapped components.

\subsubsection{Changes in the sampling rate} 
The changes in the sampling rate of encrypted signals can imply a change in the entropy and dimension (therefore, in security) of the used key. Considering again the use of modular functions of the form $f(z) = 1 + z^n$ with $n$ power of $2$, changes in the sampling rate which are powers of $2$ can be implemented  in the encrypted domain following the procedure we explain in the next two paragraphs. 

\paragraph*{Upsampling} 
For the upsampling, we only need to perform a change of variable in the polynomial ring. For an upsampling $x[l/G]$ with $G > 1$ in the ring $R_q[z]$, we just replace $z'= z^{G}$, ending in $R_q[z^{G}]$. Hence, for upsampling we apply $x(z^{G})$ and consider $f(z^{G})$ and the secret key $s(z^{G})$. Since the variables of our polynomial rings can only involve natural degrees, $G$ has to be a natural number. In our case, using $f(z) = 1 + z^n$ with $n$ power of $2$, $G$ must also be a power of $2$. After increasing the number of samples, the encrypted signal can then be low-pass filtered through a homomorphic convolution, obtaining the encryption of the interpolated signal. Regarding security, the change of variable implies an increase in the lattice dimension; however, the entropy of the key remains unchanged. From the point of view of the key owner, the key is the same, simply considering different degrees for the coefficients.

\paragraph*{Downsampling} 
Considering the ring $R_q[z]$, in order to perform a downsampling, we apply a change of variable $z^{1/G}$ with $1/G < 1$. If the corresponding coefficients of the polynomials from the ring $R_q[z^{1/G}]$ with no integer degree are discarded (plain decimation), we end up with the ciphertext $\left( c_0(z^{1/G}), c_1(z^{1/G}) \right)$ and secret key $s(z^{1/G})$. As in the case of upsampling, it is considered that $G$ is a power of $2$. 
Back in the variable $z$, decrypting $\bm{c} = (c_0, c_1)$ implies calculating $c_0 + c_1s$, where $s$ is the secret key. Hence, a downsampling of the encrypted message involves performing a decimation of both $c_0$ and the result of the multiplication of $c_1$ with $s$. 
After an upsampling with a factor $G$, we can directly perform the corresponding downsampling by $G$ without any impact on the number of ring elements that form the ciphertext. In contrast, for downsampling without relying on a previous upsampling, we need to use the polyphase decomposition of the decryption circuit, with the particularity that we are working in a ring with negacyclic convolutions instead of the typical cyclic convolutions. 
Therefore, the downsampling of the ciphertext $\bm{c} = (c_0, c_1)$ by a factor $G$ is equivalent to the first component of the polyphase decomposition in $G$ components of $\bm{c}$. If decryption computes $c_0 + c_1s$, the decryption of the decimated ciphertext would compute
\begin{equation*}
c'_0(z) + c^{(0)}\left( z \right)s^{(0)}\left( z \right) + z\sum_{i = 1}^{G - 1} c^{(i)}\left( z \right)s^{(G - i)}\left( z \right) \bmod{1 + z^{\frac{n}{G}}},
\end{equation*}
where $c'_0(z)$ is the downsampling of $c_0(z)$ and both $c^{(i)}\left( z \right)$ and $s^{(i)}\left( z \right)$ are the $i$-th polyphase components of $c_1$ and $s$, respectively. 
Now, we can reduce the ciphertext to a function of a single key. For this purpose, we can use $G-1$ concatenated relinearizations with the corresponding key homomorphisms $h_i^{(j)}$ with $i = 0, \dots, \lceil \log_t q \rceil -1$ and $j = 1, \ldots, G - 1$, which can be performed in just one step (see Section~\ref{sec:relinearization}). 
Regarding security, the entropy and size of the key are reduced in proportion to the downsampling factor.

\subsubsection{Reflection} 
We denote the reflection of the signal $a$ by $a^{\textit{ref}}$.
As the ciphertext $\bm{c}^{\textit{ref}} = (c_0^{\textit{ref}}, c_1^{\textit{ref}})$ contains the reflection of the encrypted signal under the key $s^{\textit{ref}}$, to implement the reflection of the ciphertext $\bm{c} = (c_0, c_1)$ we have to use a key change of $s^{\textit{ref}}$ instead of $s$.
Finally, the key change circuit can be represented as a relinearization of the decryption circuit $c_0^{\textit{ref}} + 0s + c_1^{\textit{ref}}s^{\textit{ref}}$, therefore considering $c_1 = 0$ in the decryption circuit introduced in Section~\ref{sec:relinearization}.

\subsubsection{Modulation and demodulation}
\label{modulation}
Typical modulations involve the multiplication by a periodic carrier. This element-wise multiplication can be addressed by the general method presented in Section~\ref{element-wise-processing}. However, the element-wise multiplication between a ciphertext and a known periodic carrier can be efficiently implemented through the method proposed in Section~\ref{periodic-element-wise-processing} which takes advantage of the periodic structure of the carrier signal, and achieves better efficiency by reducing the size of the needed relinearization matrices.

\subsection{Encrypted Matrix Multiplication}
\label{encrypted_matrix_multiplication}
We rely on Yagle's \cite{Yagle95} method to write a matrix multiplication as a single polynomial product to implement matrix multiplications on RLWE-based encrypted signals.

For calculating a matrix multiplication of size $N \times N$ as $\bm{C} = \bm{A}\bm{B}$, Yagle proposes to compute $c(z) = a(z)b(z)$, where we will denote the elements of matrices or polynomial coefficients with two or one subscripts respectively, such that 
\begin{align*} 
c(z) =& \sum_{i = 0}^{N^3 + N^2 - N - 1} c_iz^i,\;
a(z) = \sum_{i = 0}^{N-1} \sum_{j = 0}^{N-1} a_{i + jN} z^{i + jN},\\
b(z) =& \sum_{i = 0}^{N-1} \sum_{j = 0}^{N-1} b_{N-1-i + jN} z^{N (N - 1 - i + jN)}, 
\end{align*} 
with $a_{i, j} = a_{i + jN}$, $b_{i, j} = b_{N-1-i + jN} $, $c_{i, j} = c_ {N^2 - N + i + jN^2}$ and integers $i$ and $j$ such that $0 \leq i, j < N$. 

This imposes a lower bound $n \geq N^3 + N^2 - N - 1$ on the needed maximum degree $n$ of the modular function $f(z) = 1 + z^n$, in order to store the result of a matrix multiplication of size $N \times N$ within the cryptosystem. 
Possible applications of this encrypted matrix multiplication algorithms for signal processing comprise, among others, linear codes and computing $N$ encrypted linear transforms of $N$ different signals of size $N$ through a single polynomial product. 

\paragraph{Operations with the scaling variable $d$} 
The necessary lattice size for encrypting Yagle's matrix product in an encrypted algorithm can be very large, so the method can become computationally too expensive even for matrices of moderate sizes. Yagle's approach uses a scaling variable (denoted $s$ in~\cite{Yagle95}) to lower the number of coefficients of the proposed polynomials, reducing also the dimension of the considered lattice. This comes at the cost of imposing certain conditions on the magnitude of the elements of the result. 

Assuming a plaintext from a ring $R_t[z] = Z_t[z] / f(z)$ with $f(z) = 1 + z^n$ and $n$ power of $2$, we can reduce the degree of the polynomial of the modular function and still get the desired result. For this, we need to do a change of variable $z = dw$ and a change of the modular function by $f(w) = 1 + w^l$, $1 \leq l < n$, being $l$ the desired new degree (due to the cryptosystem requirements, $l$ must be a power of two). 
Therefore, if all the elements of the resulting matrix $\bm{C}$ are less than $d^l$, they can be recovered from the result with a base-$d$ decomposition of all the coefficients. For more details on using the scaling variable $d$ we refer the reader to~\cite{Yagle95}. 

This approach can be adapted to the operations described in this work in order to achieve a size reduction of the polynomials. Nevertheless, its use makes it difficult to perform subsequent encrypted homomorphic operations.

\subsection{Encrypted CRC (Cyclic Redundancy Check)} 
Given a generator polynomial $g(z)$ of maximum degree $n - k$, and a message $m(z)$ of maximum degree $k-1$,  CCCs (Cyclic Convolution Codes) encode the signal $m(z)$ as the polynomial product $m(z) g(z)$. 
After encrypting signals $g(z)$ and $m(z)$, their polynomial product can be homomorphically performed if the result fits in the length allowed by the ring $R_t[z]$, i.e., it does not \emph{wrap around}. 

Consequently, these types of codes seem to perfectly adapt to the structure of RLWE cryptosystems, enabling the application of new encrypted operations, such as encrypted CRC checks of the encrypted message. 
Some specific types of cyclic convolution codes, such as BCH or Reed-Solomon, require the use of the NTT and INTT for encoding the messages. The calculation of the encrypted NTT and INTT has been addressed in Section~\ref{sec:generalizeETs}.
Below, we include an example of the use of cyclic codes for the reduction of the cipher expansion in RLWE or $m$-RLWE based cryptosystems.

\paragraph{Cyclic codes for better cipher expansion}
In~\cite{PG14}, the authors show that in practical situations, if the least significant bits of the encrypted coefficients are discarded, the decryption error rate does not increase significantly. This line of thought can also be found in other recent works~\cite{Brakerski12} showing how discarding the least significant bits does not increase too much the ciphertext noise in a scale-invariant cryptosystem.

Therefore, a possible improvement would be the homomorphic application of a cyclic code to the encrypted values, in such a way that we could discard more bits and protect against the quantization errors without decoding first. Of course, there exists a trade-off between the increase of the polynomial size (due to the introduced redundancy in the message) and the number of discarded bits in the polynomial coefficients.

Fortunately, when the messages have a size $k$ smaller than $n$ we can apply the cyclic encoding without increasing the polynomial size of the ciphertexts, therefore achieving a reduction in the cipher expansion.
Regarding the increase of computational cost at decryption, the key owner only has to apply the corresponding cyclic decoding after the decryption of the encoded message.

\subsection{Generic Linear Transforms for encrypted vectors}
\label{sec:genericLinearTransforms}
We can implement any kind of linear transform by relying on the method presented in Section~\ref{element-wise-processing} to perform the element-wise multiplication between a ciphertext and a cleartext.
In matrix form, the element-wise multiplication can be seen as a multiplication between the ciphertext components and a diagonal matrix whose diagonal is composed of the cleartext coefficients. 
We briefly introduce a generalization of the previously considered diagonal matrix to a general square matrix. With this approach, we can perform any linear transform of an encrypted signal, provided that the matrix considered for the linear transform is available in cleartext.

For completeness, we show below the process for calculating the product between the public matrix and the encrypted vector. Additionally, we also exemplify its use for a typical signal processing application: interleaving.

\subsubsection{Implementation of the Linear Transform for an encrypted vector}
We follow an analogous process to Section~\ref{element-wise-processing}. First, we consider the ciphertext $\bm{c} = (c_0, c_1)$, whose decryption circuit is $c_0 + c_1s$, and the linear transform represented by the square matrix $\bm{L}$ of size $n \times n$. 
If we denote the polynomial $c'_0$ as the result of the multiplication $\bm{L}\bm{c_0}$, the decryption circuit in matrix form will be $\bm{c}'_0 + \bm{L} \bm{C_1} \bm{s}$.

Additionally, the product $\bm{L} \bm{C}_1 \bm{s}$ can be expressed equivalently in polynomial form as $\sum_{j = 0}^{n-1} c^{(j)}(z) s_j$, being $c^{(j)}(z)$ the polynomial whose coefficients are the $j$-th column of the matrix product $\bm{L} \bm{C}_1$. Consequently, the new decryption circuit has the form $c'_0(z) + \sum_{j = 0}^{n-1} c^{(j)}(z)s_j$.

Finally, considering $n$ key homomorphisms $h_i^{(j)}$ with $i = 0, \ldots, \lceil \log_t q \rceil - 1$ and $j = 0, \ldots, n-1$, in which $h_i^{(j)}$ has the coefficient $s_j$ ``encrypted'' under the key $s$, we can do a unique relinearization by concatenating all the $h_i^{(j)}$ and the corresponding polynomials $c^{(j)}(z)$.

Regarding the computational cost, the only difference with respect to the element-wise multiplication between a ciphertext and a cleartext shown in Section \ref{element-wise-processing} is the following: instead of multiplying the coefficients of the ciphertext with a diagonal matrix, here we use a general square matrix; that is, if we have $N/n$ ciphertexts (where $N \geq n$), then we perform: a) $\mathcal{O}(N)$ products between coefficients for the element-wise case, and b) $\mathcal{O}(Nn)$ products with the na\"{i}ve matrix multiplication algorithm between a square matrix and a vector (we are computing $N/n$ matrix multiplications among vectors and matrices with size $n$ and $n \times n$ respectively).

In any case, the part of the algorithm that determines the total computation time is the execution of $N/n$ relinearization steps with $2n \lceil \log_t{q} \rceil$ polynomial products each; i.e., approximately $\mathcal{O}(Nn{(\log_2{n})}^2)$ products between coefficients. As the relinearization process is the same for both cases, the computational cost for the general (known) linear transform of an encrypted vector is approximately the same as the suggested method for element-wise product between a ciphertext and a cleartext shown in Section~\ref{element-wise-processing}.

\subsubsection{Interleaving}
The interleaving process can be represented as a matrix product with a concatenation of permutation matrices, which conform one ``interleaving matrix''; therefore, we can implement the interleaving of the encrypted signal as a linear transform. As an example, this can be useful when performing an encrypted matrix multiplication (see Section~\ref{encrypted_matrix_multiplication}), because by changing some rows of the interleaving matrices for other rows which contain all zeros, we can relocate the coefficients of the result and zero those coefficients which are not needed.

Hence, the computational cost for the relinearization processes involved in our interleaving is the same as for the case of linear transforms discussed in Section~\ref{sec:genericLinearTransforms}. Additionally, the remaining cost of encrypted interleaving is smaller than the cost of an encrypted linear transform, as the interleaving prior to relinearization is much faster than a matrix product.

\section{Conclusions}
We have presented a novel way of using Number Theoretic Transforms paired with lattice-based cryptosystems to take advantage of the polynomial structure of typical signal processing applications and enable a wide range of unattended secure signal processing primitives for noninteractive privacy-preserving processing of sensitive signals.

On the one hand, we show a parameterization of RLWE-based cryptosystems to fully optimize the use of underlying NTTs to speed up polynomial products; additionally we show how to perform cyclic, negacyclic and generalized convolutions in the encrypted domain, encrypted component-wise products, and efficient encrypted NTT, either by applying pre- and post-processing operations, or in a fully unattended manner through the use of relinearization primitives.

We illustrate the use of our proposed approaches in several composable signal processing blocks, ranging from generalized convolutions to error correcting codes and matrix-based operations. Therefore, this work opens up a wide variety of novel secure signal processing primitives over fully encrypted signals in a non-interactive way, working either with polynomial or component-wise operations, and efficiently batching SIMD processes.

\bibliographystyle{IEEEtran}
\bibliography{bibliografia}

\end{document}